n%\documentclass[twocolumn,preprintnumbers,superscriptaddress,nofootinbib,aps,prd,floatfix]{revtex4}
\documentclass[twocolumn,preprintnumbers,groupedaddress,nofootinbib,aps,prd,floatfix]{revtex4}

\pdfoutput=1

\usepackage{amsmath,amssymb}
\usepackage{graphicx} %loads the graphicx.sty package 
\usepackage{epstopdf} %loads the epstopdf.sty package 
\usepackage{slashed}
\usepackage{color}
\usepackage{multirow}
\usepackage{footmisc}
\usepackage{hyperref}
\usepackage{subfigure}
\usepackage{snapshot}

\usepackage{array}
\newcolumntype{L}[1]{>{\raggedright\let\newline\\\arraybackslash\hspace{0pt}}m{#1}}
\newcolumntype{C}[1]{>{\centering\let\newline\\\arraybackslash\hspace{0pt}}m{#1}}
\newcolumntype{R}[1]{>{\raggedleft\let\newline\\\arraybackslash\hspace{0pt}}m{#1}}

\DefineFNsymbols*{lamport}{\dagger\ddagger\S\P\|{**}{\dagger\dagger}{\ddagger\ddagger}}

\renewcommand{\thefootnote}{\fnsymbol{footnote}}

\renewcommand{\d}{{\mathrm{d}}}
\newcommand{\be}{\begin{eqnarray*}}
\newcommand{\ee}{\end{eqnarray*}}

\newcommand{\bee}{\begin{eqnarray}}
\newcommand{\eee}{\end{eqnarray}}
\newcommand{\beeq}{\begin{equation}}
\newcommand{\eeeq}{\end{equation}}

\renewcommand{\vec}{\bf}

\begin{document}

%%%%%%%%%%%%%%%%%%%%%%%%%%%%%%%%%%%%%%%%%%%%%%%%%%%%%%%%%%
\title{Discovering Higgs boson pair production through rare final
  states at a 100 TeV collider}
%%%%%%%%%%%%%%%%%%%%%%%%%%%%%%%%%%%%%%%%%%%%%%%%%%%%%%%%%%
%
%

\begin{abstract}
We consider the process of Higgs boson pair production at a
  future proton collider with centre-of-mass energy of 100~TeV,
  focusing on rare final states that include a bottom-anti-bottom
  quark pair and multiple isolated leptons: $hh \rightarrow (b\bar{b}) + n \ell + X$, $n = \{2,4\}$, $X = \{ \slashed{E}_T, \gamma, -\}$. We construct experimental search strategies for observing
the process through these channels and make suggestions on the desired requirements for the detector design of the future collider.
\end{abstract}
%%%%%%%%%%%%%%%%%%%%%%%%%%%%%%%%%%%%%%%%%%%%%%%%%%%%%%%%%%

\author{Andreas Papaefstathiou} 
\email{apapaefs@cern.ch}
\affiliation{PH Department, TH Group, CERN,\\ CH-1211 Geneva 23, Switzerland}

%\pacs{}
\preprint{CERN-PH-TH-2015-088, MCnet-15-08}

%\twocolumn[{
%\begin{@twocolumnfalse}
\maketitle
%\end{@twocolumnfalse}
%\flushbottom
%}]

%%%%%%%%%%%%%%%%%%%%%%%%%%%%%%%%%%%%%%%%%%%%%%%%%%%%%%%%%%
\section{Introduction}
\label{sec:intro}
%%%%%%%%%%%%%%%%%%%%%%%%%%%%%%%%%%%%%%%%5
In the `aftermath' of the Large Hadron Collider's (LHC) `Run 1' at 7 and
8 TeV, we are left with the grim possibility that striking new effects
will be beyond the reach of the collider, perhaps even in `Run 2' at
13~TeV. The LHC will continue to collect data into a high-luminosity `Run 3' phase, possibly at 14~TeV, which will allow us to perform precision tests on the
properties of the Higgs boson and the underlying mechanism of
electroweak symmetry breaking (EWSB). Taking a rather pessimistic point of
view when it comes to beyond-the-Standard Model physics, a future
circular hadron collider (FCC-hh), at a potential proton-proton
centre-of-mass energy of 100~TeV, will resume this task, zooming into
the properties of the Higgs boson. An important process that has
received considerable attention in the recent years, as part of this
effort, is that of Higgs boson pair production. Examining this
process, several aspects of EWSB can be probed: e.g. the consistency
of the self-coupling of the Higgs boson with the Standard Model
expectation~\cite{Baur:2002qd, Baur:2003gp, Dolan:2012rv,
  Baglio:2012np, Barr:2013tda, Dolan:2013rja, Papaefstathiou:2012qe,
  Goertz:2013kp, Goertz:2013eka,
  Maierhofer:2013sha,Englert:2014uqa,Liu:2014rva}, whether it comes as
part of a doublet or is affected by new physics
effects~\cite{Contino:2010mh, Dolan:2012ac, Craig:2013hca,
  Gupta:2013zza, Killick:2013mya, Choi:2013qra, Cao:2013si,
  Nhung:2013lpa, Galloway:2013dma, Ellwanger:2013ova, Han:2013sga,
  No:2013wsa, McCullough:2013rea, Grober:2010yv, Contino:2012xk,
  Gillioz:2012se, Kribs:2012kz, Dawson:2012mk, Chen:2014xwa,
  Nishiwaki:2013cma, Liu:2013woa, Enkhbat:2013oba, Heng:2013cya,
  Frederix:2014hta, Baglio:2014nea, Hespel:2014sla,
  Bhattacherjee:2014bca, Liu:2014rba, Cao:2014kya, Maltoni:2014eza,
  Martin-Lozano:2015dja, vanBeekveld:2015tka}, as well as effects of higher-dimensional operators~\cite{Goertz:2014qta, Azatov:2015oxa, Goertz:2015dba}. 

The $hh$ process is currently under investigation at the LHC, both
by phenomenologists and experimentalists. The ATLAS and CMS collaborations have already
presented, as a ``warm-up'' exercise, results for resonant and non-resonant
production of $hh$ using `Run 1' results~\cite{Aad:2014yja,
  CMS-PAS-HIG-13-032, Khachatryan:2015yea}. Detection of this process remains, however, a serious challenge, even at the end of the high-luminosity run of the LHC, where 3~ab$^{-1}$ of integrated luminosity will be collected. The question examined in this article is whether a 100 TeV proton-proton collider can yield significant contributions to the investigation of this process. 

At present, it is clear that the task will not be trivial, even at
the FCC-hh, but initial results encourage optimism. Several studies
examining the $hh\rightarrow (b\bar{b}) (\gamma\gamma)$ channel, with diverse assumptions on detector performance, indicate that it would
provide a clear signal at the end of a 3~ab$^{-1}$ run of the
FCC-hh~\cite{Barr:2014sga, Azatov:2015oxa}. This process falls into
the `rare but clean' category, allowing good rejection of relatively
manageable backgrounds with high signal efficiency, on account of the
prospect of excellent resolution in the photon momentum
measurement. An obvious question is whether the increase of
energy opens up access to other such channels, previously unavailable
at the LHC due to the small total cross section. In the present
article we consider a simple set of such channels, where one Higgs
boson is allowed to decay via $h\rightarrow b\bar{b}$ and the other to
either a pair of gauge bosons $h \rightarrow ZZ, Z\gamma, W^+W^-$,
with subsequent decay to leptons, or directly to leptons $h
\rightarrow (\tau^+\tau^-) / (\mu^+ \mu^-)$. We will not consider
hadronic decays of the $\tau$ lepton, as these will require
reconsideration of tagging algorithms versus large QCD backgrounds
that will rely crucially on the, presently unknown, detector performance
of the future collider.\footnote{We note also the recent study at the LHC at 14 and the FCC-hh at 100~TeV of the three-lepton final state arising from $hh \rightarrow (W^+W^-) (W^+W^-)$ that appeared in~\cite{Li:2015yia}.} 

The final states considered, their branching ratios and cross sections
at 14~TeV are shown in Table~\ref{tab:fs}. We also give the relevant
numbers for the $hh \rightarrow (b\bar{b})(\gamma\gamma)$ process,
even though it has not been considered here. Evidently, of the
multilepton channels examined here, all but the $hh
\to (b\bar{b}) (W^+ W^-)$ channel provide a negligible number of
events at the LHC at 14~TeV. These calculations, and the rest of the signal
cross sections in this article, are based on the EFT-approximate total
NNLO cross section for Higgs boson pair production of $\sigma(14~\mathrm{TeV}) = 40.2$~fb and
$\sigma(100~\mathrm{TeV}) =
1638$~fb~\cite{deFlorian:2013jea}.\footnote{For further details on the
  cross section calculations see also~\cite{Glover:1987nx,
    Dawson:1998py, Djouadi:1999rca, Plehn:1996wb, deFlorian:2013uza,
    Grigo:2013rya, Maltoni:2014eza}.} At the FCC-hh, the factor $\sim 40$
increase in cross section, with respect to the LHC at 14~TeV, allows for exploration of these rare channels,
given that they provide a non-negligible inclusive event sample. In
what follows and the rest of this article, we define $\ell = \{e,
\mu\}$. 

It is important to mention that
the theoretical uncertainties on both the signal cross section and
differential distributions are at present rather large. These include
$\mathcal{O}(10\%)$ due to scale uncertainty, $\mathcal{O}(10\%)$ due
to parton density function uncertainty~\cite{deFlorian:2013jea, deFlorian:2015moa} and an estimated
$\mathcal{O}(10\%)$ due to the unknown finite top mass effects at
next-to-leading order in QCD~\cite{Maltoni:2014eza}. At present, the exact electroweak
corrections at next-to-leading order also remain uncertain, estimated
to be $\mathcal{O}(5\%)$ for the case of single Higgs boson production
at the LHC~\cite{Actis:2008ug}. We expect that higher-order
predictions for the signal as well as the backgrounds, both QCD and
electroweak, at next-to-leading order and beyond, will
be substantially improved in time for the construction and data taking
of a future hadron collider, which is foreseen to take place in 20-30 years, and we will neglect them in this initial
study for the sake of simplicity. Moreover, the current estimate of
$\mathcal{O}(30\%)$ on the total theoretical uncertainty are not expected
to substantially alter the conclusions of the analyses.

The paper is organised as follows: we first give details of the
simulation of detector effects and the generation of Monte Carlo event
samples used for the analyses. Subsequently, we go through each final
state, highlighting the important attributes and potential for
measurement at the FCC-hh. The main results, in Tables~\ref{tb:bb4l_bkg}, ~\ref{tb:bb2lg_bkg},~\ref{tb:bb2ell_bkg}
and~\ref{tb:bb2mu_bkg} appear as expected number of events at 3~ab$^{-1}$ of integrated luminosity and can be trivially translated into cross sections or
events at higher luminosities. We conclude by summarizing and making general
recommendations on desired features of the FCC-hh detectors.

\begin{table*}[!t]
\begin{ruledtabular}
\begin{tabular}{llll}
channel & BR & $\sigma(14~\mathrm{TeV})$~(fb) & $\sigma(100~\mathrm{TeV})$~(fb)\\\hline
$hh \to   (b\bar{b}) (ZZ) \to (b\bar{b}) (\ell^+ \ell^- \ell^{'+} \ell^{'-})$ & 0.016\% & 0.006  & 0.26 \\
$hh \to (b\bar{b}) (Z \gamma) \to (b\bar{b}) (\ell^+ \ell^- \gamma)$ & 0.013\%  & 0.005  & 0.21 \\
$hh \to (b\bar{b}) (W^+ W^-) \to (b\bar{b}) (\ell^+ \ell'^- + \slashed{E})$ & 1.658\% &  0.667 & 27.16 \\
$hh \to (b\bar{b}) (\tau^+ \tau^-) \to (b\bar{b}) (\ell^+ \ell'^- + \slashed{E})$ & 0.893\% &  0.360 & 14.63  \\
$hh \to (b\bar{b}) (\mu^+ \mu^-) $ &  0.025\% & 0.010 & 0.42 \\\hline
$hh \to (b\bar{b}) (\gamma \gamma) $ &  0.263\% & 0.106 & 4.31\\
\end{tabular}
\label{tab:fs}
\caption{Higgs boson pair production rare final state branching ratios (BR)
  and cross sections ($\sigma$, in femtobarn) that are considered in the present article. Decays to $\tau$ leptons that subsequently decay
  to electrons or muons are included. The $hh\rightarrow (b\bar{b}) (\gamma\gamma)$ channel is not considered here and given for comparison.}
\end{ruledtabular}
\end{table*}

%%%%%%%%%%%%%%%%%%%%%%%%%%%%%%%%%%%%%%
\section{Event generation and detector simulation}
%%%%%%%%%%%%%%%%%%%%%%%%%%%%%%%%%%%%%%
\subsection{Detector simulation}
In what follows, we consider all particles within a pseudorapidity of
$|\eta| < 5$ and $p_T > 500$~MeV. Following~\cite{Barr:2014sga}, we smear the momenta of all reconstructed objects according to~\cite{TheATLAScollaboration:performance2} for jets and muons and
for electrons according
to~\cite{TheATLAScollaboration:performance1}. We also apply the
relevant reconstruction efficiencies according
to~\cite{TheATLAScollaboration:performance2}  for jets and muons
and~\cite{TheATLAScollaboration:performance1} for electrons. We simulate $b$-jet tagging by looking for jets containing $B$-hadrons and applying a flat b-tagging efficiency of 70\% and a
mis-tag rate of 1\% for light-flavour jets. We do not consider
$c$-jets. We do not apply any smearing to the missing transverse
energy. 

We reconstruct jets using the anti-$k_t$ algorithm available in the
\texttt{FastJet} package~\cite{Cacciari:2011ma, Cacciari:2005hq}, with a radius
parameter of $R=0.4$. We only consider jets with $p_T > 40$~GeV within
$|\eta| < 3$ in our analysis. The jet-to-lepton mis-identification probability is
taken to be $\mathcal{P}_{j\rightarrow \ell} = 0.0048 \times
 \mathrm{e}^{-0.035 p_{Tj}/\mathrm{GeV}}$, as
 in~\cite{Barr:2014sga}. We apply a transverse momentum cut on all leptons of $p_T > 20$~GeV and require them to lie within a pseudorapidity $|\eta| < 2.5$. We also consider the mis-tagging of two bottom
quarks with a flat probability of 1\% for each mis-tag, corresponding to a $b$-jet
identification rate of 70\% and demand that they lie within $|\eta| <
2.5$. We demand all leptons to be isolated, an isolated lepton having $\sum_i p_{T,i}$
less than 15\% of its transverse momentum in a cone of $\Delta R =
0.2$ around it. 

Since the design of the detectors for FCC-hh is still under
development, we consider in conjunction to the above, which we call
`LHC parametrization' of the detector effects, an alternative `ideal'
parametrization. This is obtained by setting all efficiencies to 100\%, within
the given considered acceptance regions for jets and leptons, and by removing all momentum smearing effects. Table~\ref{tab:detparam} summarizes
the differences between the two parametrizations. The mis-tagging rates
for $b$-jets, leptons and photon were kept identical in both
parametrizations. We note, however, that these were not found to be
particularly important for the channels considered here, provided that they remain at the levels that can be achieved at the high-luminosity LHC.

\begin{table}[!t]
\begin{ruledtabular}
\begin{tabular}{lll}
property & LHC~param. & ideal~param. \\\hline
$p_{\mathrm{b-tag}}$ & 70\% & 80\% \\
$\epsilon(j)$ & Ref.~\cite{TheATLAScollaboration:performance2} & 100\% \\
$\epsilon(e)$ & Ref.~\cite{TheATLAScollaboration:performance1} & 100\%\\
$\epsilon(\mu)$ & Ref.~\cite{TheATLAScollaboration:performance2}  & 100\%\\
$\sigma(j)$ & Ref.~\cite{TheATLAScollaboration:performance2} & 0 \\
$\sigma(e)$ &  Ref.~\cite{TheATLAScollaboration:performance1}  & 0 \\
$\sigma(\mu)$ & Ref.~\cite{TheATLAScollaboration:performance2} & 0 \\
\end{tabular}
\end{ruledtabular}
\caption{The differences between the two detector parametrizations we
  consider. One is effectively an `ideal' detector, with perfect
  efficiency within the considered pseudorapidity ranges for each type
  of object, whereas the other is an LHC-like parametrisation, with
  equivalent assumptions as for the high-luminosity LHC. The $b$-tagging probability, $p_{\mathrm{b-tag}}$, the efficiency for jets, electrons and muons ($\epsilon(j/e/\mu)$) and the corresponding standard deviation of the smearing Gaussian are given ($\sigma(j/e/\mu)$). }
\label{tab:detparam}
\end{table}
\subsection{Event generation}

We generate the signal at leading order using the \texttt{Herwig++}
event generator~\cite{Bahr:2008pv, Gieseke:2011na, Arnold:2012fq,
  Bellm:2013lba} interfaced with the \texttt{OpenLoops} package for
the one-loop amplitudes~\cite{Cascioli:2011va,
  Maierhofer:2013sha}.\footnote{Note that $\mathcal{O}(10\%)$
  systematic uncertainties can be introduced by using the showered LO sample
  instead of a sample that includes higher-order real emission matrix
  elements. An extensive discussion of the parton shower uncertainties
  in Higgs boson pair production can be found
  in~\cite{Maierhofer:2013sha}. For the
  purposes of this initial study, it is sufficient to use LO $hh$
  production.} The backgrounds have been generated with the
\texttt{MadGraph 5/aMC@NLO} package~\cite{Frixione:2010ra,
  Frederix:2011zi, Alwall:2014hca}, at next-to-leading order (NLO) in
QCD, except for the case of the $t\bar{t}$ background, which was
generated at leading order and merged with the \texttt{Herwig++}
parton shower using the MLM algorithm, including $t\bar{t} + 1$~parton
matrix elements. For the latter, we normalized the cross section to
the total NLO cross section. All simulations include modelling of
hadronization as well as the underlying event through multiple parton
interactions, as they are made available in \texttt{Herwig++}. No
simulation of additional interacting protons (pile-up) is included in
this study. The \texttt{CT10nlo} parton density
function~\cite{Lai:2010vv} set was used for all parts of the simulation. 

To facilitate the Monte Carlo event generation, we reduce the rather large $t\bar{t}$ cross section at 100~TeV by applying the following generation-level cuts to the final state objects $(\ell^+ b \nu_\ell) (\ell'^-\bar{b} \bar{\nu}_{\ell'})$:\footnote{We have
checked that a full MC@NLO $t\bar{t}$ sample produces
similar efficiencies before running out of statistics.}
\begin{eqnarray}\label{eq:ttcuts}
&p_{T,b} > 40~\mathrm{GeV},~p_{T,\ell} > 30~\mathrm{GeV},\nonumber \\
&|\eta_\ell| < 2.5,\nonumber \\
&0.1 < \Delta (b,b) < 2.0, 0.1 < \Delta (\ell,\ell) < 2.0.
\end{eqnarray}
We also apply generation-level cuts to $b$-quarks in other
backgrounds. These are indicated on the corresponding tables.
%%%%%%%%%%%%%%%%%%%%%%%%%%%%%%%%%%%%%%
\section{Analysis}
%%%%%%%%%%%%%%%%%%%%%%%%%%%%%%%%%%%%%%

%%%%%%%%%%%%%%%%%%%%%%%%%%%%%%%%%%%%%%%%%%%%%%%%%%%%%%5
\subsection{$hh \to (b \bar{b}) (4\ell)$}
\label{sec:analysis1}
%%%%%%%%%%%%%%%%%%%%%%%%%%%%%%%%%%%%%%%%%%%%%%%%%%%%%%%%
Since the $h\rightarrow ZZ\rightarrow
4\ell$ channel has played an important role in the discovery of the
Higgs boson, it is reasonable to ask whether the four-lepton final state, in association with $h\rightarrow (b\bar{b})$, could be
employed equivalently in the di-Higgs channel. Unfortunately, at
the LHC at 14~TeV, the SM cross section in this channel is hopelessly
low: $\sigma((b\bar{b}) + 4\ell)_{14~\mathrm{TeV}} \sim 6\times10^{-3}$~fb, giving an
expected number of $\sim 20$ events at the end of the high-luminosity run at
3~ab$^{-1}$. Hence, even considering just triggering and basic
acceptance cuts, one can conclude that this channel will never be
observed at the LHC. At a 100~TeV collider, the cross section
increases to about 0.26~fb, leading to $\sim 780$ events at 3~ab$^{-1}$. Evidently,
the channel is still challenging, even at this future collider. Nevertheless, the final state is easier to reconstruct than
others, and one should consider whether significance could in
principle be obtained. Particularly interesting is the scenario of an
integrated luminosity of 30~ab$^{-1}$, where one would have an initial sample of $7800$ events. 

\begin{figure*}[h]
  \centering
   \subfigure[]{\includegraphics[width=0.31\textwidth]{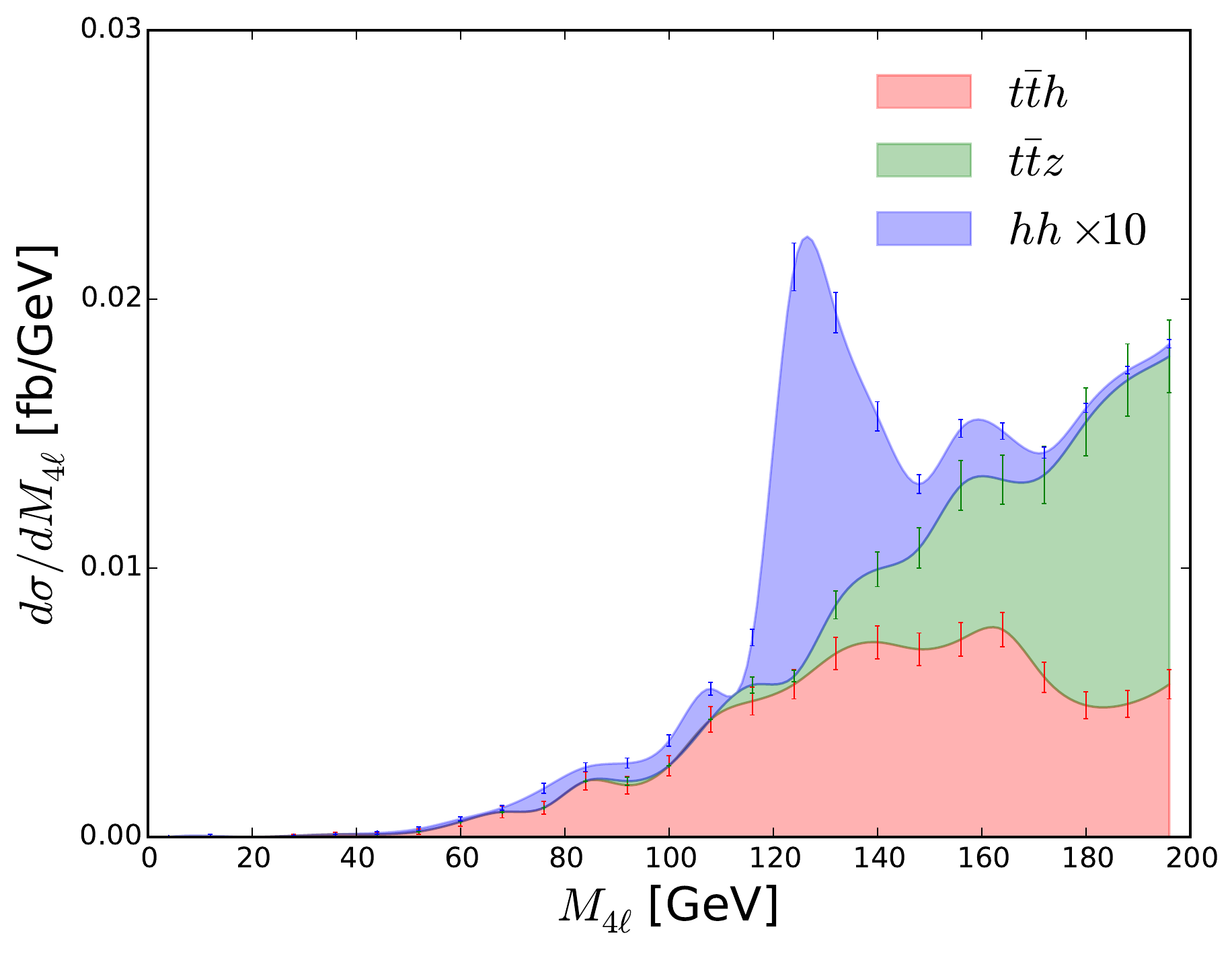}}
  \hfill
   \subfigure[]{\includegraphics[width=0.31\textwidth]{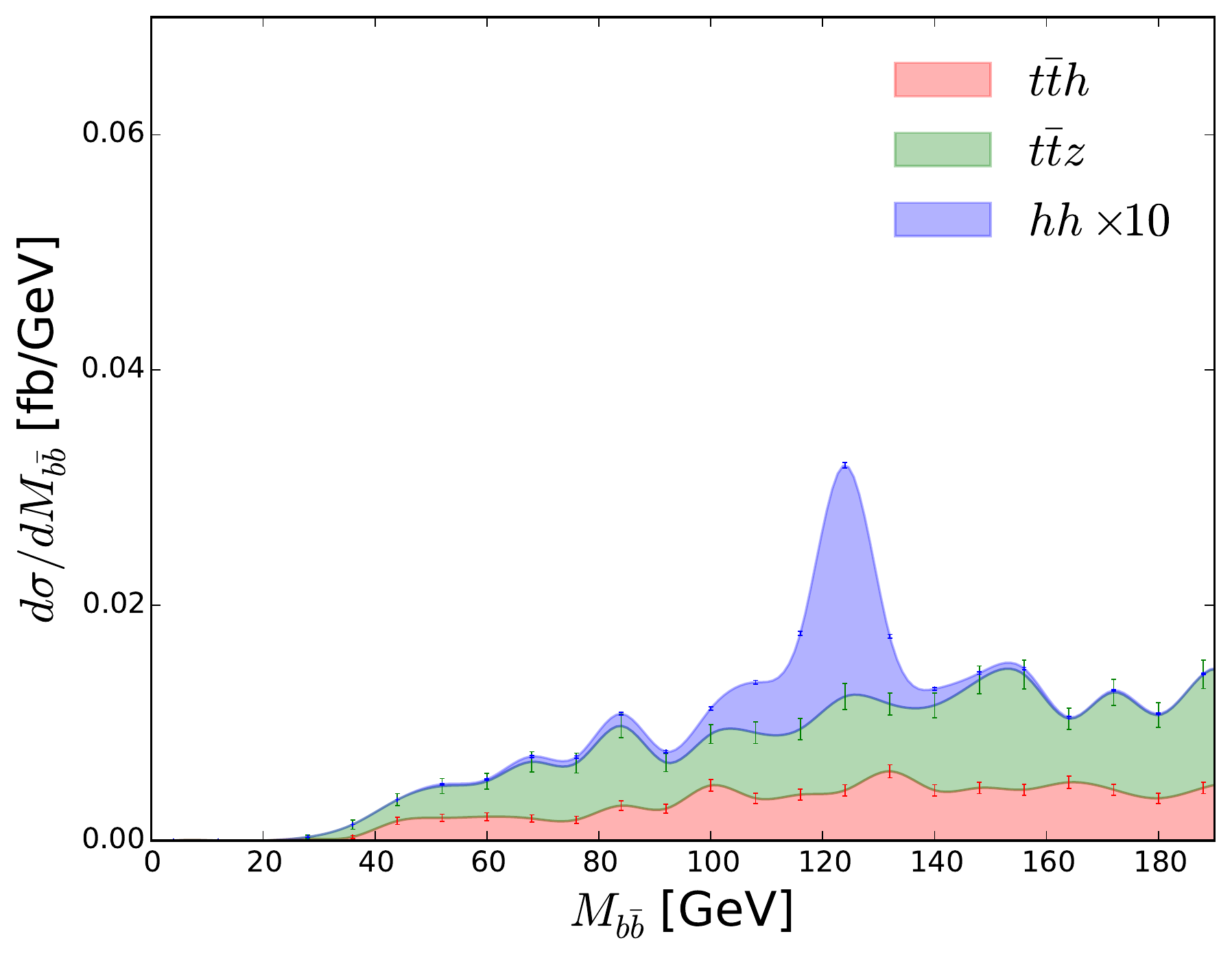}}
  \hfill
   \subfigure[]{\includegraphics[width=0.325\textwidth]{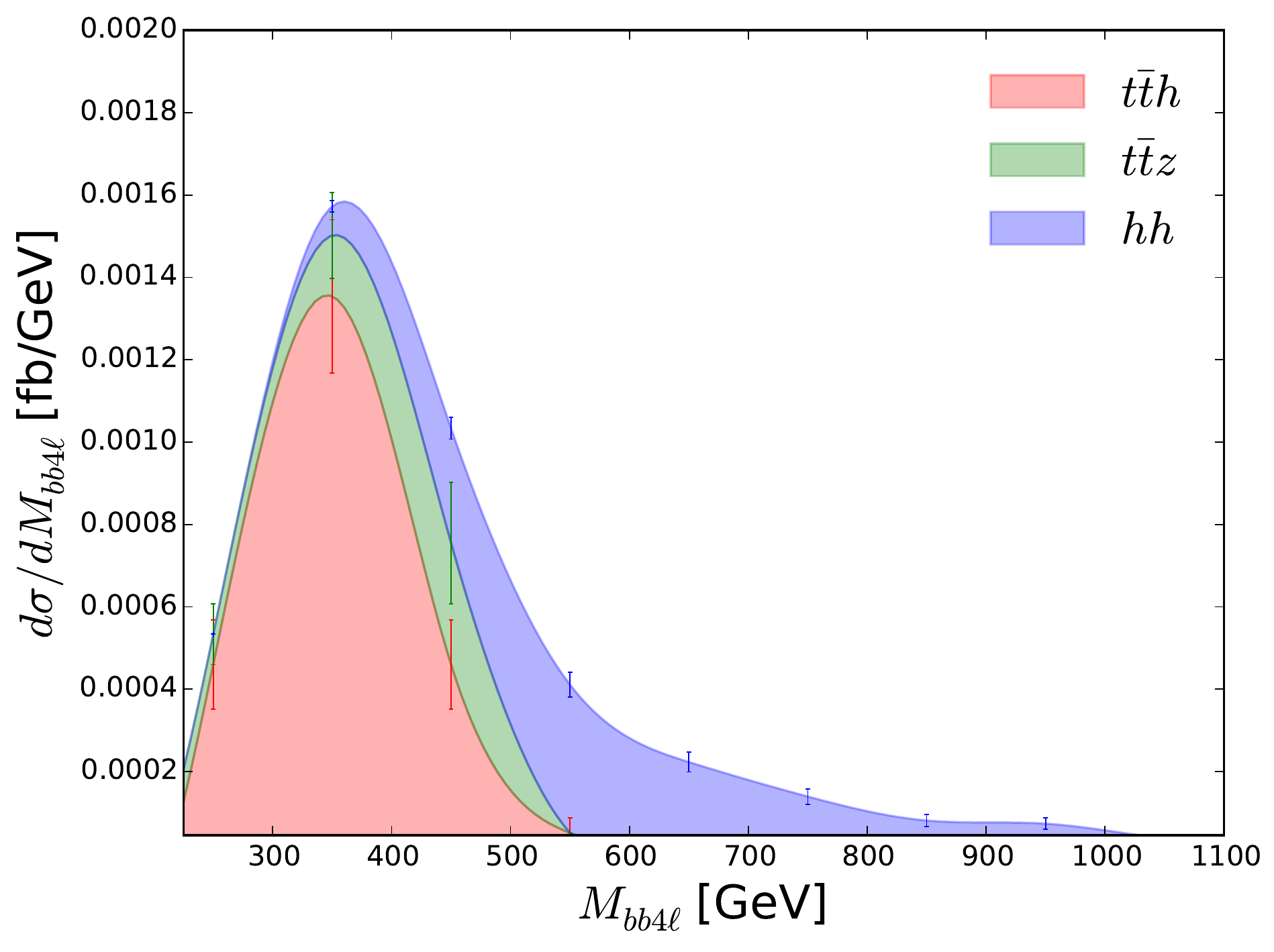}}
  \caption{\label{fig:bbZZ} The final reconstructed observables for
    the $(b\bar{b})(4\ell)$ final state, for the LHC parametrization: the invariant mass of the
four-leptons, the invariant mass of the $b$-jet pair and the invariant
mass of all six reconstructed objects. In the former two, the signal
has been multiplied by a factor of 10 for clarity. The invariant mass of
all objects is given after
cuts on the prior two observables have been imposed. The Monte Carlo statistical errors
are shown for each data sample to help distinguish statistical fluctuations from physical
features. }
\end{figure*}

The backgrounds relevant to this process are listed in Table~\ref{tb:bb4l_bkg}. Here, we only
consider mis-tagging of a single lepton, with the the dominant process
in this case being $W^\pm Zh$. Evidently, there are 6 objects relevant
to the hard process: two $b$-jets and 4 leptons. We demand pairs of
leptons of opposite charge and same flavour as well as two identified $b$-jets. As a simulation of a
possible 4-lepton trigger, we ask for the following staggered lepton
cuts: $p_{T,\ell _{\{1,2,3,4\}}} > \{ 35, 30, 25, 20 \}$~GeV. Since the
signal in this case is not expected to possess a large amount of
missing transverse energy, we impose $\slashed{E}_T < 100$~GeV. It was
observed that the distance between all leptons in the $hh$ signal is
substantially smaller than in most of the background processes, and
hence a cut of $\Delta R ( \ell _1, \ell_ j ) < 1.0$, with $j = \{2,3,4\}$,
is imposed. We apply cuts
tailored to rejecting events with two on-shell $Z$ bosons: if there
are two combinations of same-flavour opposite-sign leptons that have
an invariant mass in $m_{\ell^+\ell^-} \in (80, 100)$~GeV, we reject
the event. We also demand that no single pair of same-flavour
opposite-sign leptons possesses a mass above 120~GeV. The final
reconstructed observables are the invariant mass of the
four-leptons, the invariant mass of the $b$-jet pair and the invariant
mass of all six reconstructed objects. These observables are shown for the $hh$ signal and the two significant backgrounds, $t\bar{t}h$ and $t\bar{t}Z$, after
the aforementioned cuts in Fig.~\ref{fig:bbZZ}. To obtain the final result we further impose the following cuts:
\begin{eqnarray}
M_{bb} \in (100, 150)~\mathrm{GeV},~ M_{4\ell} \in (110,140)~\mathrm{GeV},
\end{eqnarray} 
and no cut on the invariant mass of all the reconstructed objects. 
\begin{table*}[h]
\begin{ruledtabular}
\begin{tabular}{llll}
channel & $\sigma(100~\mathrm{TeV})$~(fb) &  $N_{3~\mathrm{ab}^{-1}}(\mathrm{cuts,~ideal})$ &$N_{3~\mathrm{ab}^{-1}}(\mathrm{cuts,~LHC})$\\\hline
$\mathbf{hh} \rightarrow (b\bar{b})  (\ell^+ \ell^- \ell^{'+} \ell^{'-})$ &  0.26 & $13.0^{+0.3}_{-0.3}$ & $4.1^{+0.2}_{-0.2}$ \\\hline
$\mathbf{t\bar{t} h} \rightarrow (\ell^+ b \nu_\ell) (\ell'^- \bar{b} \bar{\nu}_{\ell'}) (2 \ell) $ & 193.6 & $30.4^{+1.4}_{-1.4}$ & $10.9^{+0.7}_{-0.7}$ \\
$\mathbf{t\bar{t} Z} \rightarrow (\ell^+ b \nu_\ell) (\ell'^- \bar{b}
  \bar{\nu}_{\ell'}) (2 \ell) $ &  256.7 & $6.6^{+2.8}_{-1.5}$ & $2.5^{+1.7}_{-0.7}$ \\
$\mathbf{Zh} \rightarrow (b\bar{b})(4 \ell)$ &  2.29  &
                                                        $\mathcal{O}(10^{-1})$ & $\mathcal{O}(10^{-1})$ \\
$\mathbf{ZZZ} \rightarrow (4 \ell)(b\bar{b}) $ & 0.53 & $\mathcal{O}(10^{-1})$  & $\mathcal{O}(10^{-1})$  \\
$\mathbf{b\bar{b} h} \rightarrow b\bar{b}(4 \ell)$, $p_{T,b} > 15$~GeV
        &  0.26  & $\mathcal{O}(1)$  & $\mathcal{O}(10^{-1})$ \\
$\mathbf{ZZh} \rightarrow (4 \ell)(b\bar{b}) $ & 0.12& $\mathcal{O}(10^{-3})$  & $\mathcal{O}(10^{-3})$ \\\hline
$\mathbf{ZZ} \rightarrow (4 \ell)$+mis-tagged $b\bar{b}$ & 781.4& $\mathcal{O}(10^{-2})$  & $\mathcal{O}(10^{-2})$  \\
$\mathbf{hZ} \rightarrow (4 \ell)$+mis-tagged $b\bar{b}$ & 68.2& $\mathcal{O}(10^{-3})$  & $\mathcal{O}(10^{-3})$  \\
$\mathbf{W\pm ZZ} \rightarrow (\ell \nu_\ell) (\ell^+\ell^-) (b\bar{b})$ + mis-tagged $\ell$ &  7.5 & $\mathcal{O}(10^{-2})$  &  $\mathcal{O}(10^{-2})$ \\
$\mathbf{W^\pm Zh} \rightarrow (\ell \nu_\ell) (\ell^+\ell^-) (b\bar{b})$ +mis-tagged $\ell$ & 1.4 & $\mathcal{O}(10^{-2})$  &$\mathcal{O}(10^{-3})$  \\
\end{tabular}
\end{ruledtabular}
\caption{The signal and backgrounds that have been considered for the $
  (b\bar{b})(\ell^+ \ell^- \ell^{'+} \ell^{'-})$ channel are shown in
  the second column. All the background cross sections have been
  calculated at NLO using the \texttt{MadGraph5/aMC@NLO} package. We
  show the resulting number of expected events,
  $N_{3~\mathrm{ab}^{-1}}$, at an integrated luminosity of 3~ab$^{-1}$
  for the two detector parametrizations in the third and fourth
  columns. We show 1$\sigma$-equivalent errors, derived according to the Poisson distribution.}
\label{tb:bb4l_bkg}
\end{table*}

\begin{figure*}[!htb]
  \centering
   \subfigure[]{\includegraphics[width=0.45\textwidth]{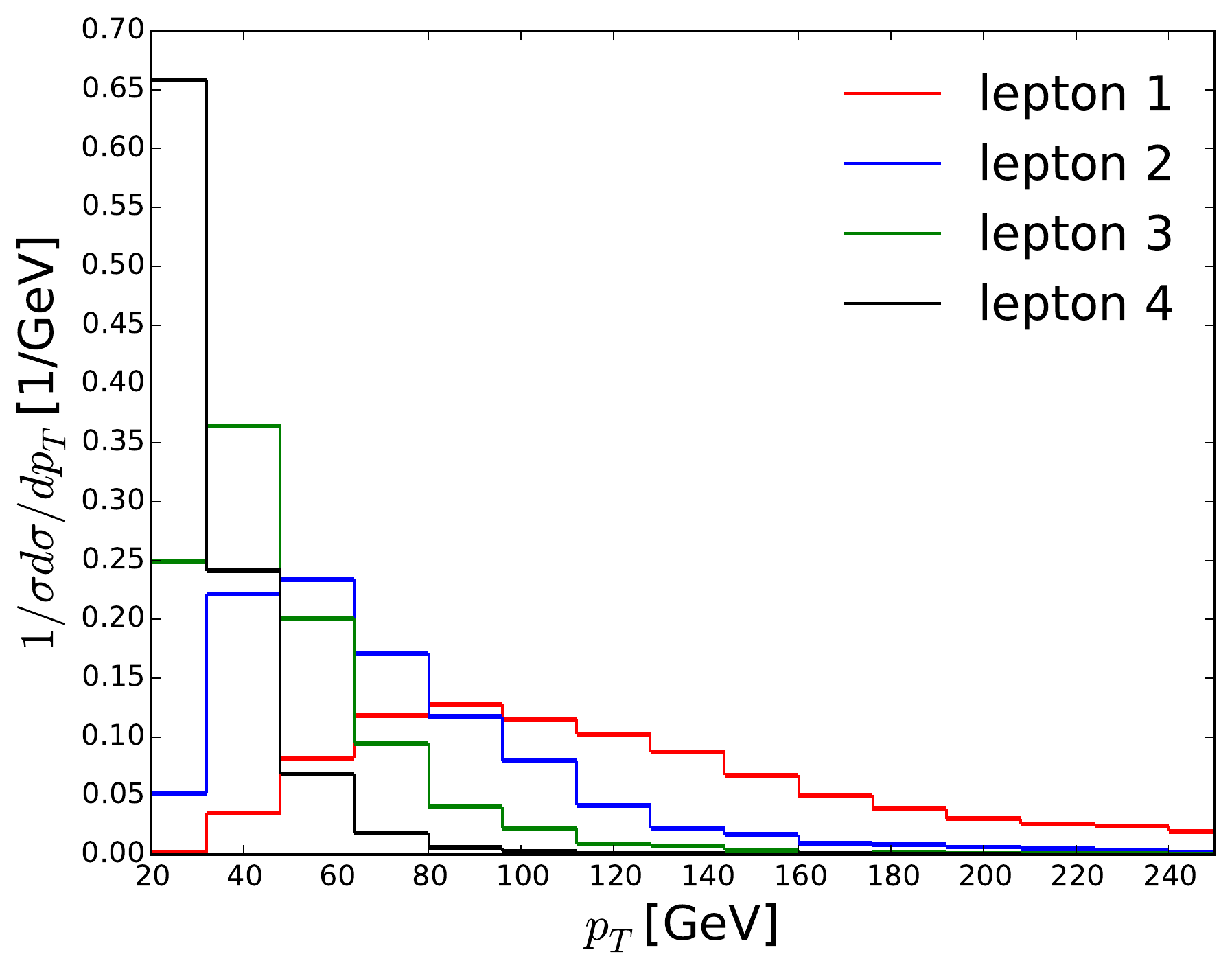}}
\hfill
   \subfigure[]{\includegraphics[width=0.45\textwidth]{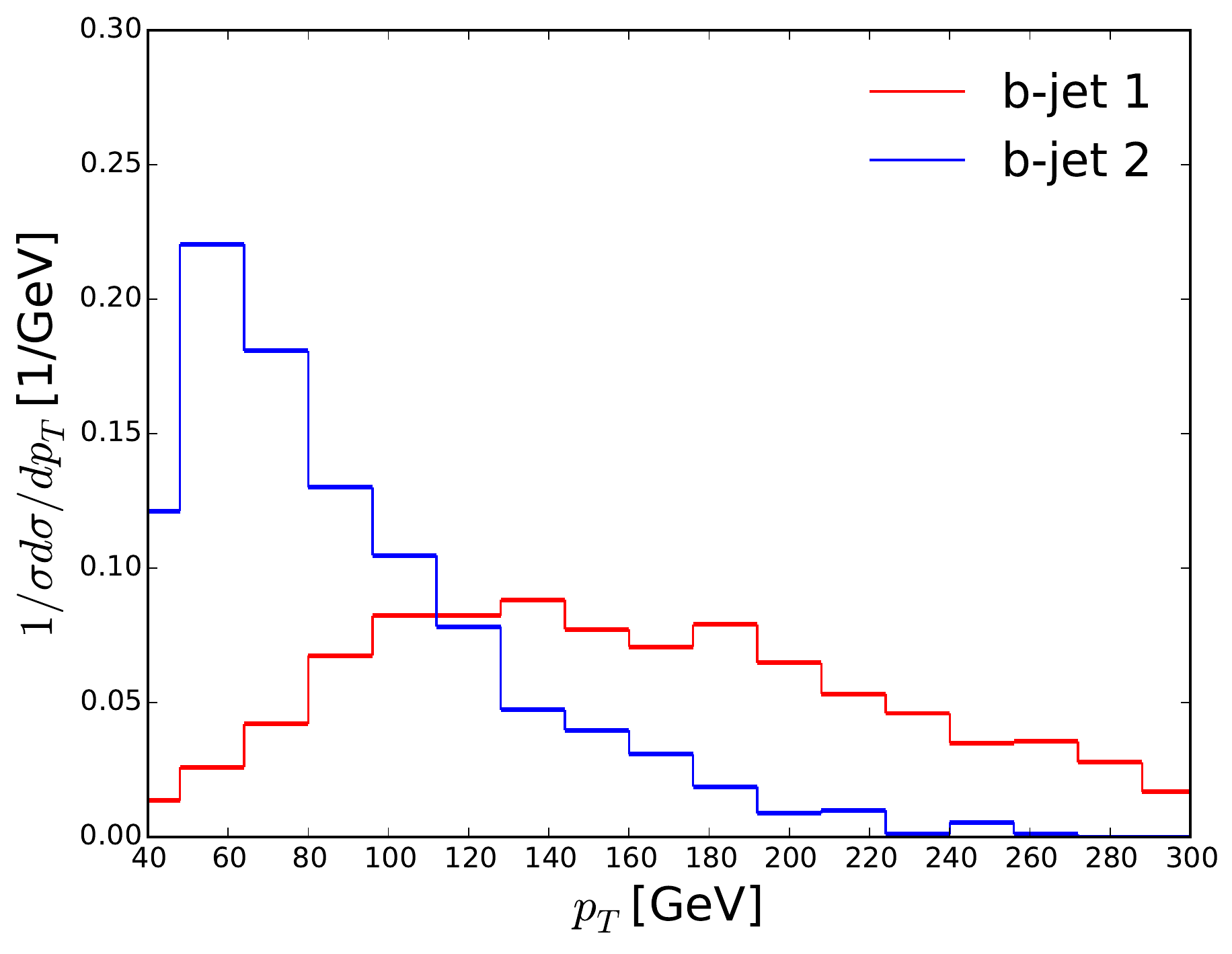}}
  \caption{\label{fig:bbZZ_pt4ell} The transverse momenta of the four leptons in the $(b\bar{b})(4\ell)$ final state prior to any cuts are shown on the left, for the $hh$ signal. The leptons are ordered in $p_T$ with $p_{T,\ell i} > p_{T,\ell j}$ for $j > i$. The right-hand figure shows the ordered transverse momenta of the $b$-jets for completeness. }
\end{figure*}

It is important to emphasize at this point a crucial element of this
analysis, namely the minimal cuts that should be applied each of the
four leptons in the final state to avoid excessive signal
rejection. In particular, for the fourth, softest, lepton in the final
state, about $65 \%$ of the events fall in
$20~\mathrm{GeV} < p_{T,\ell 4} < 30$~GeV. Not including this bin in $p_{T,\ell 4}$ would automatically reduce the
signal by more than a factor of two. This is demonstrated in Fig.~\ref{fig:bbZZ_pt4ell}, where the ordered transverse momenta of the four leptons are shown. The ordered transverse momenta of the $b$-jets are shown for completeness.

%%%%%%%%%%%%%%%%%%%%%%%%%%%%%%%%%%%%%%
\subsection{$hh \to (\bar{b}b) (\ell^+ \ell^- \gamma)$}
\label{sec:analysis2}
%%%%%%%%%%%%%%%%%%%%%%%%%%%%%%%%%%%%%%
The $hh \to (\bar{b}b) (\ell^+ \ell^- \gamma)$ channel contains two
leptons from the on-shell $Z$ decay and a hard photon, and possesses a
cross section only slightly lower than $hh \to (b \bar{b})
(4\ell)$. The backgrounds are, however, substantially larger in this
case. Here we only include the most significant irreducible ones,
coming from ${b\bar{b} Z\gamma}$, $t\bar{t}\gamma$, and $hZ \gamma$, as well as the
dominant reducible ones, where a photon is mis-tagged in $b\bar{b} Z$
or $t\bar{t}$ production.

In the analysis of this channel we require two leptons of the same
flavour with $p_{T,\ell_{\{1,2\}}} > \{ 40, 35\}$~GeV, two anti-$k_t$ $R=0.4$ $b$-jets with $p_{T,b
  _{\{1,2\}}} > \{ 60, 40\}$~GeV, $\slashed{E}_T < 80$~GeV and a
photon with $p_{T,\gamma} > 40$~GeV. No isolation requirements are
imposed on the photon. We ask for
$\Delta R ( \ell _1, \ell_ 2 ) < 1.8$, $\Delta R ( \ell _1, \gamma) <
1.5$ and $ 0.5 < \Delta R ( b_1, b_2 ) < 2.0$. We construct the
invariant mass of the $b$-jet pair and the invariant mass of the
two-lepton and photon system and impose the cuts: $100 < M_{b\bar{b}}
< 150$~GeV and $100 < M_{\ell^+ \ell^- \gamma} < 150$~GeV.

Even after these cuts, the $b\bar{b} Z\gamma$ background dominates the
resulting sample, resulting to a signal-to-background ratio of
$\mathcal{O}(2-3\%)$ with only $\mathcal{O}(10)$ signal events at
3~ab$^{-1}$ of integrated luminosity. Therefore, this channel is not
expected to provide significant information at the 100~TeV pp collider, unless a significant
alteration of the $hh$ channel is manifest due to new physics effects. 
\begin{table*}[!t]
\begin{ruledtabular}
\begin{tabular}{llll}
channel & $\sigma(100~\mathrm{TeV})$~(fb) &  $N_{3~\mathrm{ab}^{-1}}(\mathrm{cuts,~ideal})$ &$N_{3~\mathrm{ab}^{-1}}(\mathrm{cuts,~LHC})$\\\hline
$\mathbf{hh} \rightarrow (b\bar{b})  (\ell^+ \ell^- \gamma)$ & 0.21 & $14.0^{+0.3}_{-0.3}$ & $7.5^{+0.2}_{-0.2}$ \\ \hline
$\mathbf{b\bar{b} Z\gamma} \rightarrow b\bar{b}(\ell^+ \ell^-)\gamma$,
  $p_{T,b} > 30$~GeV &  26.00$\times10^3$ & $266^{+609}_{-77}$ & $203^{+465}_{-59}$  \\
$\mathbf{t\bar{t}\gamma} \rightarrow  (L^+ b \nu_Ll) (L^-
  \bar{b} \bar{\nu}_{L}) \gamma$ & 7.94$\times10^3$ & $78^{+75}_{-23}$ & $79^{+62}_{-23}$\\
$\mathbf{hZ\gamma} \rightarrow  (b\bar{b})(\ell^+ \ell^-) \gamma$ & 1.72 & $3^{+2}_{-1}$  & $2^{+1}_{-1}$\\\hline
$\mathbf{b\bar{b}Z} \rightarrow b\bar{b} (\ell^+ \ell^-) $+mis-tagged
  $\gamma$ , $p_{T,b} > 30$~GeV  &  107.36$\times 10^3$ & $20^{+22}_{-9}$ & $21^{+16}_{-6}$ \\
$\mathbf{t\bar{t}} \rightarrow  (\ell^+ b \nu_\ell) (\ell'^-
  \bar{b} \bar{\nu}_{\ell'})$+mis-tagged
  $\gamma$, cuts as in Eq.~\ref{eq:ttcuts} & 25.08$\times10^3$ & $14^{+2}_{-2}$ & $10^{+1}_{-1}$
%$\mathbf{b\bar{b}h} \rightarrow b\bar{b} (\ell^+ \ell^-) $+mis-tagged
 % $\gamma$ , $p_{T,b} > 30$~GeV  &  26.81  & &\\
%$\mathbf{ZW^\pm  \gamma} \rightarrow (b\bar{b}) (\ell^\pm \nu_\ell)$+mis-tagged
 % $\ell$ & 0.49 & & \\
%$\mathbf{hW^\pm  \gamma} \rightarrow (b\bar{b}) (\ell^\pm\nu_\ell)$+mis-tagged $\ell$ & 0.04& & d\\
%\midrule
\end{tabular}
\end{ruledtabular}
\caption{The signal and backgrounds that have been considered in the $(b\bar{b})
  (\ell^+ \ell^- \gamma)$ channel. All the background cross sections
  are given at NLO using the \texttt{MadGraph5/aMC@NLO} package.  The $t\bar{t}$ process with the
    mis-tagged $\gamma$ was generated at tree-level, merged to one jet
  via the MLM method, and normalized to the NLO cross section. We show 1$\sigma$-equivalent errors, derived according to the Poisson distribution.}
\label{tb:bb2lg_bkg}
\end{table*}

%%%%%%%%%%%%%%%%%%%%%%%%%%%%%%%%%%%%%%
\subsection{$hh \to (\bar{b}b) (\ell^+ \ell^-) (+ \slashed{E}_T)$}
\label{sec:analysis3}
%%%%%%%%%%%%%%%%%%%%%%%%%%%%%%%%%%%%%%
Another interesting channel to consider is the final state that
includes a $b\bar{b}$ and two oppositely-charged leptons. This channel
receives signal contributions from three different $hh$ decay
modes. The largest contribution comes from $hh \rightarrow (b\bar{b})
(W^+ W^-)$ with the $W$s decaying (either directly, or indirectly
through taus) to electrons or muons. The second-largest contribution
to this channel comes from $hh \rightarrow (b\bar{b}) (\tau^+\tau^-)$,
with both taus decaying to electrons or muons. Both of these channels
will include final-state neutrinos, and hence will be associated with
large missing energy. The specific final state has been considered at
the LHC at 14~TeV, e.g. coming from $(b\bar{b}) (W^+ W^-)$
in.~\cite{Baglio:2012np} or included implicitly as part of the channel $(b\bar{b}) (\tau^+\tau^-)$ in~\cite{Dolan:2012rv, Baglio:2012np, Barr:2013tda, Maierhofer:2013sha}. The smallest contribution comes from $hh \rightarrow (b\bar{b}) (\mu^+\mu^-)$, i.e. through direct decay of one of the Higgs bosons to muons. This has been considered in~\cite{Baur:2003gp} at a 200 TeV proton collider, which was envisioned to have a integrated luminosity of either 600 fb$^{-1}$ or 1200 fb$^{-1}$. The channel was shown to be able to provide some information on the $hh$ process at a higher-energy collider, and hence we re-examine it here at a 100~TeV collider, allowing for the possibility of collecting a higher integrated luminosity sample. 

Due to the different origin of the leptons in the three different
signal processes, the kinematical details vary substantially. As
already mentioned, in $(b\bar{b}) (W^+ W^-)$  and $(b\bar{b})
(\tau^+\tau^-)$, we expect large missing energy. The $\tau$ leptons in
$(b\bar{b}) (\tau^+\tau^-)$ are light compared to the Higgs boson, and
hence the leptons and neutrinos in their decays are expected to be
collimated. On the other hand, in $(b\bar{b}) (W^+ W^-)$, both $W$s
are heavy, one being most of the time on-shell ($M_W \sim 80.4$~GeV)
and the other off-shell with $M_{W*}$ peaking at $\sim 40$~GeV. This implies that the distribution of the leptons and neutrinos will differ considerably from the $(b\bar{b}) (\tau^+\tau^-)$ case. This will be reflected in the analysis efficiency for the two
processes: the fact that the leptons in boosted $\tau$ decays are
collimated with the associated neutrinos makes the $h\rightarrow
\tau^+ \tau^-$ decay easier to separate from the backgrounds. The
$(b\bar{b}) (\mu^+\mu^-)$ final state is expected to have little
missing energy, possibly only due to $B$-hadron decays, and the two
muons are expected to reconstruct the Higgs boson. To account for all
of these properties, we construct two separate signal regions. One
aims to capture events containing rather large missing energy,
targeting the $(b\bar{b}) (W^+ W^-)$  and $(b\bar{b}) (\tau^+\tau^-)$
channels, whereas the second is aimed towards events with minimal
missing energy that are expected to characterize the $(b\bar{b})
(\mu^+\mu^-)$ channel. The same set of observables is considered, with
substantial variations of cuts. We consider events with two isolated
leptons, with isolation criteria as in the previous sections. We construct
their transverse momentum, $p_{T,\ell _{\{1,2\}}}$, the distance
between them, $\Delta R(\ell_1, \ell_2)$ and their invariant mass
$M_{\ell\ell}$. We ask for two $b$-jets, for which we construct the
transverse momenta, distance and invariant mass: $p_{T,b _{\{1,2\}}}$, $\Delta R(b_1,b_2)$, $M_{bb}$. We also consider the missing transverse energy, $\slashed{E}_T$ and the invariant mass of all the reconstructed objects, $M_{bb\ell\ell}$. Moreover, we consider a further observable, $M_\mathrm{reco.}$, constructed by assuming that the missing energy arising from neutrinos in the decays of the $\tau$ leptons is collinear to the observed leptons:
\begin{equation}
M_\mathrm{reco.} = \left[ p_{b_1} + p_{b_2} + (1+f_1) p_{\ell_1}  + (1 + f_2)  p_{\ell_2}\right]^2 \;,
\end{equation}
where $p_{b_i}$, $p_{\ell_i}$ are the observed momenta of the $i$-th
$b$-jet and $i$-th lepton and $f_{1,2}$ are constants of
proportionality between the neutrino and lepton momenta from the decay
of the two $\tau$ leptons: $p_{\nu_i} = f_i p_{\ell_i}$. These can be calculated from the observed
missing transverse energy by inverting the missing transverse momentum
balance relation $L \mathbf{f} = \mathbf{\slashed{E}}$, where $L$ is
the matrix $L_i^j = p_{{\ell_i}}^j$, for which the superscript denotes
the component of the $i$-th lepton momentum, $j = \{x,y\}$ and $E$ and
$\mathbf{f}$ are the vectors $\mathbf{\slashed{E}} = ( \slashed{E}^x,
\slashed{E}^y )$ and $\mathbf{f} = (f_1, f_2)$. We consider this
observable for the $(b\bar{b}) (W^+ W^-)$ signal sample as well, even
though the collinearity approximation fails, as it is still expected
to be correlated with the invariant mass of the Higgs boson pair. One
may also define the reconstructed Higgs boson mass from the
reconstructed $\tau$ lepton momenta: $M_\mathrm{h, reco.} = \left[ (1+f_1) p_{\ell_1}  + (1 + f_2)  p_{\ell_2}\right]^2$. 

We call the signal regions SR$_{\slashed{E}}$ and SR$_\mu$, corresponding to the $hh$ decay modes $(b\bar{b}) (W^+ W^-)$, $(b\bar{b}) (\tau^+\tau^-)$ and $(b\bar{b}) (\mu^+\mu^-)$ respectively. Table~\ref{tb:srs} shows the cuts chosen for these observables for these three signal regions. Note that the $M_{T2}$ observable can also be constructed for the rejection of the $t\bar{t}$ background, but we do not take this approach here~\cite{Barr:2013tda}.

The backgrounds considered for the $(\bar{b}b) (\ell^+ \ell^-) (+ \slashed{E}_T)$ final state include the irreducible ones coming from $t\bar{t}$, with subsequent semi-leptonic decays of both top quarks, those from $b\bar{b}Z$ with decays of the $Z$ boson to leptons, those from $b\bar{b}h$ with subsequent decays of the Higgs boson to two leptons, as well as the resonant $hZ$ and $ZZ$ backgrounds. We also consider here the mis-tagging of a jet to a single lepton through the $b\bar{b}W^\pm$ channel and the mis-tagging of $b\bar{b}$ in the $\ell^+ \ell^-$+jets background, which was considered using $\ell^+ \ell^-$+1 parton at NLO. As before, we do not consider mis-identification of $c$-jets to $b$-jets. 

We show the resulting events after analysis for the signal region
SR$_\slashed{E}$ in Table~\ref{tb:bb2ell_bkg}. Both the ``LHC'' and the
``ideal'' parametrizations are shown. Evidently, since we are using
the same set of cuts for both scenaria, the ``ideal'' parametrization
does not necessarily provide a substantial improvement to the signal
efficiency. This effect is observed in particular for the $(b\bar{b}) (W^+ W^-)$
sample. Moreover, it is evident that this channel could be addressed
by employing more advanced statistical methods: cuts could be devised
separately for the two sub-channels $(b\bar{b}) (W^+ W^-)$ and
$(b\bar{b}) (\tau^+\tau^-)$ and then combined taking into account the
correlations and cross-contamination.\footnote{Another improvement
  would involve separating the $(b\bar{b}) (\tau^+\tau^-)$ final
  states into same-flavour and different-flavour. The author thanks
  Fabio Maltoni for useful discussion on this.} For example, this could be done
via appropriate cuts on $M_\mathrm{h,reco.}$, which was found, as
expected, to peak at the Higgs boson mass for the $(b\bar{b})
(\tau^+\tau^-)$ sample and around $50$~GeV for the $(b\bar{b}) (W^+W^-)$  sample. This study is beyond the scope of this initial investigation. 

Due to the fact that both signal and backgrounds possess larger cross sections, harder cuts are imposed in this channel than in the $(b\bar{b})(4\ell)$ and $(b\bar{b})(Z\gamma)$ channels. This involves cutting on variables that are expected to have a high degree of correlation with the invariant mass of the Higgs boson pair, which implies that this channel would be less sensitive to variations of the self-coupling than, for example, the $ hh \rightarrow (b \bar{b}) (4\ell)$, where no such cuts are imposed. Nevertheless, insofar as Standard Model-like $hh$ production is concerned, this process is expected to provide important information, contributing to the detection of this channel at a 100~TeV collider, with signal-to-background ratio of $\sim 0.1$ and large statistical significance at 3 ab$^{-1}$. 

\begin{table*}[!t]
\begin{ruledtabular}
\begin{tabular}{lll}
observable & SR$_\slashed{E}$ &  SR$_\mu$ \\\hline
$\slashed{E}_T$  & $>100$~GeV & $<40$~GeV\\
$p_{T,\ell _1}$ & $>60$~GeV & $>90$~GeV \\ 
$p_{T,\ell _2}$ &$>55$~GeV  &$>60$~GeV \\ 
$\Delta R(\ell_1, \ell_2)$ & $<0.9$& $\in (1.0, 1.8)$\\ 
$M_{\ell\ell}$ & $\in (50, 80)$~GeV  &$\in (120, 130)$~GeV    \\ 
$p_{T,b_1}$ & $>90$~GeV &$>90$~GeV \\ 
$p_{T,b_2}$ & $>80$~GeV& $>80$~GeV\\ 
$\Delta R(b_1,b_2)$ &$\in (0.5, 1.3)$ & $\in (0.5, 1.5)$ \\ 
$M_{bb\ell\ell}$ & $>350$~GeV&$>350$~GeV \\
$M_{bb}$ & $\in (110,140)$~GeV& $\in (110,140)$~GeV\\
$M_\mathrm{reco.}$ & $>600$~GeV & none \\
\end{tabular}
\end{ruledtabular}
\caption{The cuts for the three signal regions constructed in the analysis of the $hh \to (\bar{b}b) (\ell^+ \ell^-) (+ \slashed{E}_T)$ channel. The signal regions SR$_\slashed{E}$ and SR$_\mu$, correspond to the $hh$ decay modes $(b\bar{b}) (W^+ W^-)$, $(b\bar{b}) (\tau^+\tau^-)$ and $(b\bar{b}) (\mu^+\mu^-)$ respectively. }
\label{tb:srs}
\end{table*}

On the other hand, the situation for the $(b\bar{b}) (\mu^+\mu^-)$ after the SR$_\mu$ cuts are applied is rather bleak: at 3~ab$^{-1}$ of integrated luminosity, only a handful of events are expected with the ``LHC'' detector parametrization with a few hundred background events, even after hard transverse momentum cuts on the muons and a tight mass window on the di-muon invariant mass around the Higgs boson mass. Because of the latter cut, turning to the ``ideal'' situation improves the signal efficiency substantially, since the smearing of the muon momenta is absent. Despite this, only $\mathcal{O}(10)$ events would be obtained at 3 ab$^{-1}$ with a similar number of background events as for the `LHC' parametrization. Hence, barring any significant enhancements of the rate due to new physics, the $(b\bar{b}) (\mu^+\mu^-)$ contribution to the $hh \to (\bar{b}b) (\ell^+ \ell^-)$ final state is not expected to provide significant information.

\begin{table*}[!t]
\begin{ruledtabular}
\begin{tabular}{llll}
channel & $\sigma(100~\mathrm{TeV})$~(fb) &  $N_{3~\mathrm{ab}^{-1}}(\mathrm{cuts,~ideal})$ &$N_{3~\mathrm{ab}^{-1}}(\mathrm{cuts,~LHC})$\\\hline
$\mathbf{hh} \rightarrow (b\bar{b}) (W^+ W^-) \rightarrow (b\bar{b}) (\ell'^+ \nu_{\ell'} \ell^- \bar{\nu}_\ell)$ & 27.16 & $20.9^{+1.8}_{-1.8}$ & $19.9^{+1.5}_{-1.5}$\\ 
$\mathbf{hh} \rightarrow (b\bar{b}) (\tau^+ \tau^-) \rightarrow (b\bar{b})  (\ell'^+ \nu_{\ell'} \bar{\nu}_\tau \ell^- \bar{\nu}_\ell\nu_\tau)$ & 14.63&$38.5^{+4.7}_{-4.7}$ & $24.3^{+3.2}_{-3.2}$ \\ 
\hline
$\mathbf{t\bar{t}} \rightarrow  (\ell^+ b \nu_\ell) (\ell'^-\bar{b} \bar{\nu}_{\ell'})$, cuts as in Eq.~\ref{eq:ttcuts} & 25.08$\times10^3$ & $34.3^{+23.2}_{-9.4}$  & $15.8^{+15.3}_{-4.8}$ \\
$\mathbf{b\bar{b} Z} \rightarrow b\bar{b}(\ell^+ \ell^-)$, $p_{T,b} >30$~GeV  &  107.36$\times 10^3$ & $257.9^{+203.7}_{-74.6}$ & $493.7^{+224.9}_{-113.4}$  \\ 
$\mathbf{Z Z} \rightarrow b\bar{b}(\ell^+ \ell^-)$& 356.0 &  $\mathcal{O}(10^{-1})$ &  $\mathcal{O}(10^{-1})$ \\ 
$\mathbf{h Z} \rightarrow b\bar{b}(\ell^+ \ell^-)$& 99.79 & $49.8^{+2.9}_{-2.9}$ & $40.4^{+2.3}_{-2.3}$    \\ 
$\mathbf{b\bar{b}h} \rightarrow b\bar{b} (\ell^+ \ell^-)$, $p_{T,b} > 30$~GeV  & 26.81 & $\mathcal{O}(1)$  &$\mathcal{O}(1)$  \\\hline
$\mathbf{b\bar{b}W^\pm }\rightarrow b\bar{b} (\ell^\pm\nu_\ell)$,
  $p_{T,b} > 30$~GeV +mis-tagged $\ell$ & 1032.6 & $\mathcal{O}(10^{-2})$  & $\mathcal{O}(10^{-2})$ \\
$\mathbf{\ell^+ \ell^-+\mathrm{jets}}\rightarrow (\ell^+ \ell^-)$ + mis-tagged $b\bar{b}$ & 2.14$\times 10^3$ &  $\mathcal{O}(10^{-2})$ & $\mathcal{O}(10^{-2})$ \\
\end{tabular}
\end{ruledtabular}
\caption{The signal and backgrounds that have been considered in the $(b\bar{b})
  (\ell^+ \ell^-+ \slashed{E})$ channel coming from $h\rightarrow W^+
  W^-$. All the background cross sections were calculated at NLO using the \texttt{MadGraph5/aMC@NLO} package. The $t\bar{t}$ process was generated at tree-level, merged to one jet
  via the MLM method, and normalized to the NLO cross section. We show 1$\sigma$-equivalent errors, derived according to the Poisson distribution.}
\label{tb:bb2ell_bkg}
\end{table*}

\begin{table*}[!t]
\begin{ruledtabular}
\begin{tabular}{llll}
channel & $\sigma(100~\mathrm{TeV})$~(fb) &  $N_{3~\mathrm{ab}^{-1}}(\mathrm{cuts,~ideal})$ &$N_{3~\mathrm{ab}^{-1}}(\mathrm{cuts,~LHC})$\\\hline
$\mathbf{hh} \rightarrow (b\bar{b})  (\mu^+ \mu^-)$ & 0.42 & $8.6^{+1.1}_{-1.1}$ & $1.8^{+0.5}_{-0.3}$\\\hline
$\mathbf{t\bar{t}} \rightarrow  (\ell^+ b \nu_\ell) (\ell'^-  \bar{b} \bar{\nu}_{\ell'})$, cuts as in Eq.~\ref{eq:ttcuts} & 25.08$\times10^3$ & $48.0^{+110.2}_{-14.0}$ &$15.8^{+15.3}_{-4.8}$  \\
$\mathbf{b\bar{b} Z} \rightarrow b\bar{b}(\ell^+ \ell^-)$, $p_{T,b} > 30$~GeV  &  107.36$\times 10^3$ & $< 73.5$ & $49.4^{+113.4}_{-14.4}$  \\ 
$\mathbf{Z Z} \rightarrow b\bar{b}(\ell^+ \ell^-)$& 356.0 & $\mathcal{O}(10^{-1})$ & $\mathcal{O}(10^{-1})$  \\ 
$\mathbf{h Z} \rightarrow b\bar{b}(\ell^+ \ell^-)$& 99.79 &$\mathcal{O}(10^{-1})$  &  $2.5^{+0.7}_{-0.4}$ \\ 
$\mathbf{b\bar{b}h} \rightarrow b\bar{b} (\ell^+ \ell^-)$, $p_{T,b} > 30$~GeV  & 26.81  &$\mathcal{O}(1)$  & $\mathcal{O}(1)$ \\\hline
$\mathbf{b\bar{b}W^\pm }\rightarrow b\bar{b} (\ell^\pm\nu_\ell)$, $p_{T,b} > 30$~GeV +mis-tagged $\ell$ & 1032.6 & $\mathcal{O}(10^{-2})$ & $\mathcal{O}(10^{-2})$ \\
$\mathbf{\ell^+ \ell^-+\mathrm{jets}}\rightarrow (\ell^+ \ell^-)$ + mis-tagged $b\bar{b}$ & 2.14$\times 10^3$ & $\mathcal{O}(10^{-2})$& $\mathcal{O}(10^{-2})$ \\
%\midrule
\end{tabular}
\end{ruledtabular}
\caption{The signal and backgrounds that have been considered in the $(b\bar{b})
  (\mu^+ \mu^-)$ channel. All the background cross sections were calculated at NLO using the
  \texttt{MadGraph5/aMC@NLO} package. The $t\bar{t}$ process was generated at tree-level, merged to one jet
  via the MLM method, and normalized to the NLO cross section. We show
  1$\sigma$-equivalent errors, derived according to the Poisson
  distribution. For the case of $b\bar{b}Z$ in the `ideal' parametrization, the '$<$' indicates the 1$\sigma$-equivalent region, since no events were obtained after analysis.}
\label{tb:bb2mu_bkg}
\end{table*}

%%%%%%%%%%%%%%%%%%%%%%%%%%%%%%%%
\section{Discussion and Conclusions}
\label{sec:conc}
%%%%%%%%%%%%%%%%%%%%%%%%%%%%%%%%
In this letter we have considered three rare, and potentially clean,
final states coming from standard model-like Higgs boson pair
production at a future 100 TeV proton-proton collider. These processes
are made available for investigation due to the fact that the total Higgs boson pair production cross section increases by a factor of $\sim 40$ over that expected at the LHC. 

For the $hh \to (b \bar{b}) (ZZ)$ channel, where both $Z$ bosons decay
to leptons, resulting in $(b \bar{b}) (4\ell)$, it was shown that a
few events could be obtained at 3~ab$^{-1}$, versus $\mathcal{O}(10)$
background events. A crucial aspect of this analysis, that should be
taken into account in detector design, is being able to observe
four-lepton final states with at least one lepton possessing transverse momentum down to $\sim 20$~GeV.

The  $hh \to (b \bar{b}) (Z\gamma)$ channel, with the $Z$ boson decaying to leptons, was also considered briefly and was found to be of negligible importance even at 30~ab$^{-1}$. This is due to large backgrounds originating from the processes $b\bar{b}Z\gamma$ and $t\bar{t}\gamma$. 

The last final state considered was that containing $(b \bar{b})$ and
2 leptons. This receives contributions from  $(b\bar{b}) (W^+ W^-)$,
$(b\bar{b}) (\tau^+\tau^-)$ and  $(b\bar{b}) (\mu^+ \mu^-)$. The
former two channels possess large missing energy due to neutrinos. The
latter channel is not expected to be associated with significant
missing energy, with the muons coming directly from the decay of the
Higgs boson. The analysis was split into two signal regions, depending
on the amount of missing energy. The signal region targeting $(b\bar{b})
(W^+ W^-)$  and $(b\bar{b}) (\tau^+\tau^-)$ with leptonic final states provides promising results, with few tens of events versus a few hundred
background events at 3~ab$^{-1}$ of integrated luminosity. This can be
further improved by designing a method that captures the individual features of the two sub-channels, as well as using flavour information for the leptons. The other signal region, designed for the $(b\bar{b}) (\mu^+ \mu^-)$ channel yields few events at 3~ab$^{-1}$, with significantly larger backgrounds. Potential enhancement of the latter channel could arise from an improvement in the resolution of the muon momenta that would allow a tighter mass window around their invariant mass.

A boost in the importance of all the channels containing a $h
\rightarrow b\bar{b}$ would similarly be obtained by improvement of
the resolution of the $b$-jet momenta, as the mass window around the
Higgs boson mass could then be shrank down to beyond the 30-50~GeV
region that has been considered here. Additionally, in a final
analysis, performed with a more complete FCC-hh detector design in
mind, boosted decision tree or neural network methods would increase
significances by taking into account the intricate correlations
between the observables, going beyond the simple ``rectangular'' cuts
that we have applied here. Moreover, a complete study should take into
account realistic estimates of the theoretical uncertainties due to
higher order corrections, as well as improved Monte Carlo descriptions
for the generation of signal samples that will become available in the
future. 

Finally, the initial investigation of the processes considered here, along with the $hh \to (b \bar{b}) (\gamma\gamma)$ process, indicate that the study of Higgs boson pair production at a future hadron collider with 100 TeV centre-of-mass energy would greatly benefit with the collection of 10 or 30~ab$^{-1}$ of integrated luminosity. 

\acknowledgments
We would like to thank Paolo Torrielli for technical assistance with
Monte Carlo event generation, Jos\'e Zurita, Florian Goertz and Li Lin
Yang for providing useful comments, as well as Nikiforos Nikiforou for useful discussions during various coffee breaks. We would also like to thank the Physics
Institute, University of Z\"urich, for allowing continuous use of their computing
resources while this project was being completed. This research is
supported by the MCnetITN FP7 Marie Curie Initial Training Network
PITN-GA-2012-315877 and a Marie Curie Intra European Fellowship within the 7th European Community Framework Programme (grant no. PIEF-GA-2013-622071).

%%%%%%%%%%%%%%%%%%%%%%%%%%%%%%%%%%%%%%%%%%%%%%%%%%%%%%%%%%

%\newpage
\bibliography{hhlept.bib}

%merlin.mbs apsrev4-1.bst 2010-07-25 4.21a (PWD, AO, DPC) hacked
%Control: key (0)
%Control: author (72) initials jnrlst
%Control: editor formatted (1) identically to author
%Control: production of article title (-1) disabled
%Control: page (0) single
%Control: year (1) truncated
%Control: production of eprint (0) enabled
\begin{thebibliography}{74}%
\makeatletter
\providecommand \@ifxundefined [1]{%
 \@ifx{#1\undefined}
}%
\providecommand \@ifnum [1]{%
 \ifnum #1\expandafter \@firstoftwo
 \else \expandafter \@secondoftwo
 \fi
}%
\providecommand \@ifx [1]{%
 \ifx #1\expandafter \@firstoftwo
 \else \expandafter \@secondoftwo
 \fi
}%
\providecommand \natexlab [1]{#1}%
\providecommand \enquote  [1]{``#1''}%
\providecommand \bibnamefont  [1]{#1}%
\providecommand \bibfnamefont [1]{#1}%
\providecommand \citenamefont [1]{#1}%
\providecommand \href@noop [0]{\@secondoftwo}%
\providecommand \href [0]{\begingroup \@sanitize@url \@href}%
\providecommand \@href[1]{\@@startlink{#1}\@@href}%
\providecommand \@@href[1]{\endgroup#1\@@endlink}%
\providecommand \@sanitize@url [0]{\catcode `\\12\catcode `\$12\catcode
  `\&12\catcode `\#12\catcode `\^12\catcode `\_12\catcode `\%12\relax}%
\providecommand \@@startlink[1]{}%
\providecommand \@@endlink[0]{}%
\providecommand \url  [0]{\begingroup\@sanitize@url \@url }%
\providecommand \@url [1]{\endgroup\@href {#1}{\urlprefix }}%
\providecommand \urlprefix  [0]{URL }%
\providecommand \Eprint [0]{\href }%
\providecommand \doibase [0]{http://dx.doi.org/}%
\providecommand \selectlanguage [0]{\@gobble}%
\providecommand \bibinfo  [0]{\@secondoftwo}%
\providecommand \bibfield  [0]{\@secondoftwo}%
\providecommand \translation [1]{[#1]}%
\providecommand \BibitemOpen [0]{}%
\providecommand \bibitemStop [0]{}%
\providecommand \bibitemNoStop [0]{.\EOS\space}%
\providecommand \EOS [0]{\spacefactor3000\relax}%
\providecommand \BibitemShut  [1]{\csname bibitem#1\endcsname}%
\let\auto@bib@innerbib\@empty
%</preamble>
\bibitem [{\citenamefont {Baur}\ \emph {et~al.}(2003)\citenamefont {Baur},
  \citenamefont {Plehn},\ and\ \citenamefont {Rainwater}}]{Baur:2002qd}%
  \BibitemOpen
  \bibfield  {author} {\bibinfo {author} {\bibfnamefont {U.}~\bibnamefont
  {Baur}}, \bibinfo {author} {\bibfnamefont {T.}~\bibnamefont {Plehn}}, \ and\
  \bibinfo {author} {\bibfnamefont {D.~L.}\ \bibnamefont {Rainwater}},\ }\href
  {\doibase 10.1103/PhysRevD.67.033003} {\bibfield  {journal} {\bibinfo
  {journal} {Phys.Rev.}\ }\textbf {\bibinfo {volume} {D67}},\ \bibinfo {pages}
  {033003} (\bibinfo {year} {2003})},\ \Eprint
  {http://arxiv.org/abs/hep-ph/0211224} {arXiv:hep-ph/0211224 [hep-ph]}
  \BibitemShut {NoStop}%
%%CITATION = HEP-PH/0211224;%%
\bibitem [{\citenamefont {Baur}\ \emph {et~al.}(2004)\citenamefont {Baur},
  \citenamefont {Plehn},\ and\ \citenamefont {Rainwater}}]{Baur:2003gp}%
  \BibitemOpen
  \bibfield  {author} {\bibinfo {author} {\bibfnamefont {U.}~\bibnamefont
  {Baur}}, \bibinfo {author} {\bibfnamefont {T.}~\bibnamefont {Plehn}}, \ and\
  \bibinfo {author} {\bibfnamefont {D.~L.}\ \bibnamefont {Rainwater}},\ }\href
  {\doibase 10.1103/PhysRevD.69.053004} {\bibfield  {journal} {\bibinfo
  {journal} {Phys.Rev.}\ }\textbf {\bibinfo {volume} {D69}},\ \bibinfo {pages}
  {053004} (\bibinfo {year} {2004})},\ \Eprint
  {http://arxiv.org/abs/hep-ph/0310056} {arXiv:hep-ph/0310056 [hep-ph]}
  \BibitemShut {NoStop}%
%%CITATION = HEP-PH/0310056;%%
\bibitem [{\citenamefont {Dolan}\ \emph {et~al.}(2012)\citenamefont {Dolan},
  \citenamefont {Englert},\ and\ \citenamefont {Spannowsky}}]{Dolan:2012rv}%
  \BibitemOpen
  \bibfield  {author} {\bibinfo {author} {\bibfnamefont {M.~J.}\ \bibnamefont
  {Dolan}}, \bibinfo {author} {\bibfnamefont {C.}~\bibnamefont {Englert}}, \
  and\ \bibinfo {author} {\bibfnamefont {M.}~\bibnamefont {Spannowsky}},\
  }\href {\doibase 10.1007/JHEP10(2012)112} {\bibfield  {journal} {\bibinfo
  {journal} {JHEP}\ }\textbf {\bibinfo {volume} {1210}},\ \bibinfo {pages}
  {112} (\bibinfo {year} {2012})},\ \Eprint {http://arxiv.org/abs/1206.5001}
  {arXiv:1206.5001 [hep-ph]} \BibitemShut {NoStop}%
%%CITATION = ARXIV:1206.5001;%%
\bibitem [{\citenamefont {Baglio}\ \emph {et~al.}(2013)\citenamefont {Baglio},
  \citenamefont {Djouadi}, \citenamefont {Gröber}, \citenamefont
  {Mühlleitner}, \citenamefont {Quevillon} \emph {et~al.}}]{Baglio:2012np}%
  \BibitemOpen
  \bibfield  {author} {\bibinfo {author} {\bibfnamefont {J.}~\bibnamefont
  {Baglio}}, \bibinfo {author} {\bibfnamefont {A.}~\bibnamefont {Djouadi}},
  \bibinfo {author} {\bibfnamefont {R.}~\bibnamefont {Gröber}}, \bibinfo
  {author} {\bibfnamefont {M.}~\bibnamefont {Mühlleitner}}, \bibinfo {author}
  {\bibfnamefont {J.}~\bibnamefont {Quevillon}},  \emph {et~al.},\ }\href
  {\doibase 10.1007/JHEP04(2013)151} {\bibfield  {journal} {\bibinfo  {journal}
  {JHEP}\ }\textbf {\bibinfo {volume} {1304}},\ \bibinfo {pages} {151}
  (\bibinfo {year} {2013})},\ \Eprint {http://arxiv.org/abs/1212.5581}
  {arXiv:1212.5581 [hep-ph]} \BibitemShut {NoStop}%
%%CITATION = ARXIV:1212.5581;%%
\bibitem [{\citenamefont {Barr}\ \emph {et~al.}(2014)\citenamefont {Barr},
  \citenamefont {Dolan}, \citenamefont {Englert},\ and\ \citenamefont
  {Spannowsky}}]{Barr:2013tda}%
  \BibitemOpen
  \bibfield  {author} {\bibinfo {author} {\bibfnamefont {A.~J.}\ \bibnamefont
  {Barr}}, \bibinfo {author} {\bibfnamefont {M.~J.}\ \bibnamefont {Dolan}},
  \bibinfo {author} {\bibfnamefont {C.}~\bibnamefont {Englert}}, \ and\
  \bibinfo {author} {\bibfnamefont {M.}~\bibnamefont {Spannowsky}},\ }\href
  {\doibase 10.1016/j.physletb.2013.12.011} {\bibfield  {journal} {\bibinfo
  {journal} {Phys.Lett.}\ }\textbf {\bibinfo {volume} {B728}},\ \bibinfo
  {pages} {308} (\bibinfo {year} {2014})},\ \Eprint
  {http://arxiv.org/abs/1309.6318} {arXiv:1309.6318 [hep-ph]} \BibitemShut
  {NoStop}%
%%CITATION = ARXIV:1309.6318;%%
\bibitem [{\citenamefont {Dolan}\ \emph {et~al.}(2014)\citenamefont {Dolan},
  \citenamefont {Englert}, \citenamefont {Greiner},\ and\ \citenamefont
  {Spannowsky}}]{Dolan:2013rja}%
  \BibitemOpen
  \bibfield  {author} {\bibinfo {author} {\bibfnamefont {M.~J.}\ \bibnamefont
  {Dolan}}, \bibinfo {author} {\bibfnamefont {C.}~\bibnamefont {Englert}},
  \bibinfo {author} {\bibfnamefont {N.}~\bibnamefont {Greiner}}, \ and\
  \bibinfo {author} {\bibfnamefont {M.}~\bibnamefont {Spannowsky}},\ }\href
  {\doibase 10.1103/PhysRevLett.112.101802} {\bibfield  {journal} {\bibinfo
  {journal} {Phys.Rev.Lett.}\ }\textbf {\bibinfo {volume} {112}},\ \bibinfo
  {pages} {101802} (\bibinfo {year} {2014})},\ \Eprint
  {http://arxiv.org/abs/1310.1084} {arXiv:1310.1084 [hep-ph]} \BibitemShut
  {NoStop}%
%%CITATION = ARXIV:1310.1084;%%
\bibitem [{\citenamefont {Papaefstathiou}\ \emph {et~al.}(2013)\citenamefont
  {Papaefstathiou}, \citenamefont {Yang},\ and\ \citenamefont
  {Zurita}}]{Papaefstathiou:2012qe}%
  \BibitemOpen
  \bibfield  {author} {\bibinfo {author} {\bibfnamefont {A.}~\bibnamefont
  {Papaefstathiou}}, \bibinfo {author} {\bibfnamefont {L.~L.}\ \bibnamefont
  {Yang}}, \ and\ \bibinfo {author} {\bibfnamefont {J.}~\bibnamefont
  {Zurita}},\ }\href {\doibase 10.1103/PhysRevD.87.011301} {\bibfield
  {journal} {\bibinfo  {journal} {Phys.Rev.}\ }\textbf {\bibinfo {volume}
  {D87}},\ \bibinfo {pages} {011301} (\bibinfo {year} {2013})},\ \Eprint
  {http://arxiv.org/abs/1209.1489} {arXiv:1209.1489 [hep-ph]} \BibitemShut
  {NoStop}%
%%CITATION = ARXIV:1209.1489;%%
\bibitem [{\citenamefont {Goertz}\ \emph
  {et~al.}(2013{\natexlab{a}})\citenamefont {Goertz}, \citenamefont
  {Papaefstathiou}, \citenamefont {Yang},\ and\ \citenamefont
  {Zurita}}]{Goertz:2013kp}%
  \BibitemOpen
  \bibfield  {author} {\bibinfo {author} {\bibfnamefont {F.}~\bibnamefont
  {Goertz}}, \bibinfo {author} {\bibfnamefont {A.}~\bibnamefont
  {Papaefstathiou}}, \bibinfo {author} {\bibfnamefont {L.~L.}\ \bibnamefont
  {Yang}}, \ and\ \bibinfo {author} {\bibfnamefont {J.}~\bibnamefont
  {Zurita}},\ }\href {\doibase 10.1007/JHEP06(2013)016} {\bibfield  {journal}
  {\bibinfo  {journal} {JHEP}\ }\textbf {\bibinfo {volume} {1306}},\ \bibinfo
  {pages} {016} (\bibinfo {year} {2013}{\natexlab{a}})},\ \Eprint
  {http://arxiv.org/abs/1301.3492} {arXiv:1301.3492 [hep-ph]} \BibitemShut
  {NoStop}%
%%CITATION = ARXIV:1301.3492;%%
\bibitem [{\citenamefont {Goertz}\ \emph
  {et~al.}(2013{\natexlab{b}})\citenamefont {Goertz}, \citenamefont
  {Papaefstathiou}, \citenamefont {Yang},\ and\ \citenamefont
  {Zurita}}]{Goertz:2013eka}%
  \BibitemOpen
  \bibfield  {author} {\bibinfo {author} {\bibfnamefont {F.}~\bibnamefont
  {Goertz}}, \bibinfo {author} {\bibfnamefont {A.}~\bibnamefont
  {Papaefstathiou}}, \bibinfo {author} {\bibfnamefont {L.~L.}\ \bibnamefont
  {Yang}}, \ and\ \bibinfo {author} {\bibfnamefont {J.}~\bibnamefont
  {Zurita}},\ }\href@noop {} {\  (\bibinfo {year} {2013}{\natexlab{b}})},\
  \Eprint {http://arxiv.org/abs/1309.3805} {arXiv:1309.3805 [hep-ph]}
  \BibitemShut {NoStop}%
%%CITATION = ARXIV:1309.3805;%%
\bibitem [{\citenamefont {Maierhoefer}\ and\ \citenamefont
  {Papaefstathiou}(2014)}]{Maierhofer:2013sha}%
  \BibitemOpen
  \bibfield  {author} {\bibinfo {author} {\bibfnamefont {P.}~\bibnamefont
  {Maierhoefer}}\ and\ \bibinfo {author} {\bibfnamefont {A.}~\bibnamefont
  {Papaefstathiou}},\ }\href {\doibase 10.1007/JHEP03(2014)126} {\bibfield
  {journal} {\bibinfo  {journal} {JHEP}\ }\textbf {\bibinfo {volume} {1403}},\
  \bibinfo {pages} {126} (\bibinfo {year} {2014})},\ \Eprint
  {http://arxiv.org/abs/1401.0007} {arXiv:1401.0007 [hep-ph]} \BibitemShut
  {NoStop}%
%%CITATION = ARXIV:1401.0007;%%
\bibitem [{\citenamefont {Englert}\ \emph {et~al.}(2015)\citenamefont
  {Englert}, \citenamefont {Krauss}, \citenamefont {Spannowsky},\ and\
  \citenamefont {Thompson}}]{Englert:2014uqa}%
  \BibitemOpen
  \bibfield  {author} {\bibinfo {author} {\bibfnamefont {C.}~\bibnamefont
  {Englert}}, \bibinfo {author} {\bibfnamefont {F.}~\bibnamefont {Krauss}},
  \bibinfo {author} {\bibfnamefont {M.}~\bibnamefont {Spannowsky}}, \ and\
  \bibinfo {author} {\bibfnamefont {J.}~\bibnamefont {Thompson}},\ }\href
  {\doibase 10.1016/j.physletb.2015.02.041} {\bibfield  {journal} {\bibinfo
  {journal} {Phys.Lett.}\ }\textbf {\bibinfo {volume} {B743}},\ \bibinfo
  {pages} {93} (\bibinfo {year} {2015})},\ \Eprint
  {http://arxiv.org/abs/1409.8074} {arXiv:1409.8074 [hep-ph]} \BibitemShut
  {NoStop}%
%%CITATION = ARXIV:1409.8074;%%
\bibitem [{\citenamefont {Liu}\ and\ \citenamefont
  {Zhang}(2014)}]{Liu:2014rva}%
  \BibitemOpen
  \bibfield  {author} {\bibinfo {author} {\bibfnamefont {T.}~\bibnamefont
  {Liu}}\ and\ \bibinfo {author} {\bibfnamefont {H.}~\bibnamefont {Zhang}},\
  }\href@noop {} {\  (\bibinfo {year} {2014})},\ \Eprint
  {http://arxiv.org/abs/1410.1855} {arXiv:1410.1855 [hep-ph]} \BibitemShut
  {NoStop}%
%%CITATION = ARXIV:1410.1855;%%
\bibitem [{\citenamefont {Contino}\ \emph {et~al.}(2010)\citenamefont
  {Contino}, \citenamefont {Grojean}, \citenamefont {Moretti}, \citenamefont
  {Piccinini},\ and\ \citenamefont {Rattazzi}}]{Contino:2010mh}%
  \BibitemOpen
  \bibfield  {author} {\bibinfo {author} {\bibfnamefont {R.}~\bibnamefont
  {Contino}}, \bibinfo {author} {\bibfnamefont {C.}~\bibnamefont {Grojean}},
  \bibinfo {author} {\bibfnamefont {M.}~\bibnamefont {Moretti}}, \bibinfo
  {author} {\bibfnamefont {F.}~\bibnamefont {Piccinini}}, \ and\ \bibinfo
  {author} {\bibfnamefont {R.}~\bibnamefont {Rattazzi}},\ }\href {\doibase
  10.1007/JHEP05(2010)089} {\bibfield  {journal} {\bibinfo  {journal} {JHEP}\
  }\textbf {\bibinfo {volume} {1005}},\ \bibinfo {pages} {089} (\bibinfo {year}
  {2010})},\ \Eprint {http://arxiv.org/abs/1002.1011} {arXiv:1002.1011
  [hep-ph]} \BibitemShut {NoStop}%
%%CITATION = ARXIV:1002.1011;%%
\bibitem [{\citenamefont {Dolan}\ \emph {et~al.}(2013)\citenamefont {Dolan},
  \citenamefont {Englert},\ and\ \citenamefont {Spannowsky}}]{Dolan:2012ac}%
  \BibitemOpen
  \bibfield  {author} {\bibinfo {author} {\bibfnamefont {M.~J.}\ \bibnamefont
  {Dolan}}, \bibinfo {author} {\bibfnamefont {C.}~\bibnamefont {Englert}}, \
  and\ \bibinfo {author} {\bibfnamefont {M.}~\bibnamefont {Spannowsky}},\
  }\href {\doibase 10.1103/PhysRevD.87.055002} {\bibfield  {journal} {\bibinfo
  {journal} {Phys.Rev.}\ }\textbf {\bibinfo {volume} {D87}},\ \bibinfo {pages}
  {055002} (\bibinfo {year} {2013})},\ \Eprint {http://arxiv.org/abs/1210.8166}
  {arXiv:1210.8166 [hep-ph]} \BibitemShut {NoStop}%
%%CITATION = ARXIV:1210.8166;%%
\bibitem [{\citenamefont {Craig}\ \emph {et~al.}(2013)\citenamefont {Craig},
  \citenamefont {Galloway},\ and\ \citenamefont {Thomas}}]{Craig:2013hca}%
  \BibitemOpen
  \bibfield  {author} {\bibinfo {author} {\bibfnamefont {N.}~\bibnamefont
  {Craig}}, \bibinfo {author} {\bibfnamefont {J.}~\bibnamefont {Galloway}}, \
  and\ \bibinfo {author} {\bibfnamefont {S.}~\bibnamefont {Thomas}},\
  }\href@noop {} {\  (\bibinfo {year} {2013})},\ \Eprint
  {http://arxiv.org/abs/1305.2424} {arXiv:1305.2424 [hep-ph]} \BibitemShut
  {NoStop}%
%%CITATION = ARXIV:1305.2424;%%
\bibitem [{\citenamefont {Gupta}\ \emph {et~al.}(2013)\citenamefont {Gupta},
  \citenamefont {Rzehak},\ and\ \citenamefont {Wells}}]{Gupta:2013zza}%
  \BibitemOpen
  \bibfield  {author} {\bibinfo {author} {\bibfnamefont {R.~S.}\ \bibnamefont
  {Gupta}}, \bibinfo {author} {\bibfnamefont {H.}~\bibnamefont {Rzehak}}, \
  and\ \bibinfo {author} {\bibfnamefont {J.~D.}\ \bibnamefont {Wells}},\ }\href
  {\doibase 10.1103/PhysRevD.88.055024} {\bibfield  {journal} {\bibinfo
  {journal} {Phys.Rev.}\ }\textbf {\bibinfo {volume} {D88}},\ \bibinfo {pages}
  {055024} (\bibinfo {year} {2013})},\ \Eprint {http://arxiv.org/abs/1305.6397}
  {arXiv:1305.6397 [hep-ph]} \BibitemShut {NoStop}%
%%CITATION = ARXIV:1305.6397;%%
\bibitem [{\citenamefont {Killick}\ \emph {et~al.}(2013)\citenamefont
  {Killick}, \citenamefont {Kumar},\ and\ \citenamefont
  {Logan}}]{Killick:2013mya}%
  \BibitemOpen
  \bibfield  {author} {\bibinfo {author} {\bibfnamefont {R.}~\bibnamefont
  {Killick}}, \bibinfo {author} {\bibfnamefont {K.}~\bibnamefont {Kumar}}, \
  and\ \bibinfo {author} {\bibfnamefont {H.~E.}\ \bibnamefont {Logan}},\ }\href
  {\doibase 10.1103/PhysRevD.88.033015} {\bibfield  {journal} {\bibinfo
  {journal} {Phys.Rev.}\ }\textbf {\bibinfo {volume} {D88}},\ \bibinfo {pages}
  {033015} (\bibinfo {year} {2013})},\ \Eprint {http://arxiv.org/abs/1305.7236}
  {arXiv:1305.7236 [hep-ph]} \BibitemShut {NoStop}%
%%CITATION = ARXIV:1305.7236;%%
\bibitem [{\citenamefont {Choi}\ \emph {et~al.}(2013)\citenamefont {Choi},
  \citenamefont {Englert},\ and\ \citenamefont {Zerwas}}]{Choi:2013qra}%
  \BibitemOpen
  \bibfield  {author} {\bibinfo {author} {\bibfnamefont {S.}~\bibnamefont
  {Choi}}, \bibinfo {author} {\bibfnamefont {C.}~\bibnamefont {Englert}}, \
  and\ \bibinfo {author} {\bibfnamefont {P.}~\bibnamefont {Zerwas}},\ }\href
  {\doibase 10.1140/epjc/s10052-013-2643-z} {\bibfield  {journal} {\bibinfo
  {journal} {Eur.Phys.J.}\ }\textbf {\bibinfo {volume} {C73}},\ \bibinfo
  {pages} {2643} (\bibinfo {year} {2013})},\ \Eprint
  {http://arxiv.org/abs/1308.5784} {arXiv:1308.5784 [hep-ph]} \BibitemShut
  {NoStop}%
%%CITATION = ARXIV:1308.5784;%%
\bibitem [{\citenamefont {Cao}\ \emph {et~al.}(2013)\citenamefont {Cao},
  \citenamefont {Heng}, \citenamefont {Shang}, \citenamefont {Wan},\ and\
  \citenamefont {Yang}}]{Cao:2013si}%
  \BibitemOpen
  \bibfield  {author} {\bibinfo {author} {\bibfnamefont {J.}~\bibnamefont
  {Cao}}, \bibinfo {author} {\bibfnamefont {Z.}~\bibnamefont {Heng}}, \bibinfo
  {author} {\bibfnamefont {L.}~\bibnamefont {Shang}}, \bibinfo {author}
  {\bibfnamefont {P.}~\bibnamefont {Wan}}, \ and\ \bibinfo {author}
  {\bibfnamefont {J.~M.}\ \bibnamefont {Yang}},\ }\href {\doibase
  10.1007/JHEP04(2013)134} {\bibfield  {journal} {\bibinfo  {journal} {JHEP}\
  }\textbf {\bibinfo {volume} {1304}},\ \bibinfo {pages} {134} (\bibinfo {year}
  {2013})},\ \Eprint {http://arxiv.org/abs/1301.6437} {arXiv:1301.6437
  [hep-ph]} \BibitemShut {NoStop}%
%%CITATION = ARXIV:1301.6437;%%
\bibitem [{\citenamefont {Nhung}\ \emph {et~al.}(2013)\citenamefont {Nhung},
  \citenamefont {Muhlleitner}, \citenamefont {Streicher},\ and\ \citenamefont
  {Walz}}]{Nhung:2013lpa}%
  \BibitemOpen
  \bibfield  {author} {\bibinfo {author} {\bibfnamefont {D.~T.}\ \bibnamefont
  {Nhung}}, \bibinfo {author} {\bibfnamefont {M.}~\bibnamefont {Muhlleitner}},
  \bibinfo {author} {\bibfnamefont {J.}~\bibnamefont {Streicher}}, \ and\
  \bibinfo {author} {\bibfnamefont {K.}~\bibnamefont {Walz}},\ }\href {\doibase
  10.1007/JHEP11(2013)181} {\bibfield  {journal} {\bibinfo  {journal} {JHEP}\
  }\textbf {\bibinfo {volume} {1311}},\ \bibinfo {pages} {181} (\bibinfo {year}
  {2013})},\ \Eprint {http://arxiv.org/abs/1306.3926} {arXiv:1306.3926
  [hep-ph]} \BibitemShut {NoStop}%
%%CITATION = ARXIV:1306.3926;%%
\bibitem [{\citenamefont {Galloway}\ \emph {et~al.}(2014)\citenamefont
  {Galloway}, \citenamefont {Luty}, \citenamefont {Tsai},\ and\ \citenamefont
  {Zhao}}]{Galloway:2013dma}%
  \BibitemOpen
  \bibfield  {author} {\bibinfo {author} {\bibfnamefont {J.}~\bibnamefont
  {Galloway}}, \bibinfo {author} {\bibfnamefont {M.~A.}\ \bibnamefont {Luty}},
  \bibinfo {author} {\bibfnamefont {Y.}~\bibnamefont {Tsai}}, \ and\ \bibinfo
  {author} {\bibfnamefont {Y.}~\bibnamefont {Zhao}},\ }\href {\doibase
  10.1103/PhysRevD.89.075003} {\bibfield  {journal} {\bibinfo  {journal}
  {Phys.Rev.}\ }\textbf {\bibinfo {volume} {D89}},\ \bibinfo {pages} {075003}
  (\bibinfo {year} {2014})},\ \Eprint {http://arxiv.org/abs/1306.6354}
  {arXiv:1306.6354 [hep-ph]} \BibitemShut {NoStop}%
%%CITATION = ARXIV:1306.6354;%%
\bibitem [{\citenamefont {Ellwanger}(2013)}]{Ellwanger:2013ova}%
  \BibitemOpen
  \bibfield  {author} {\bibinfo {author} {\bibfnamefont {U.}~\bibnamefont
  {Ellwanger}},\ }\href {\doibase 10.1007/JHEP08(2013)077} {\bibfield
  {journal} {\bibinfo  {journal} {JHEP}\ }\textbf {\bibinfo {volume} {1308}},\
  \bibinfo {pages} {077} (\bibinfo {year} {2013})},\ \Eprint
  {http://arxiv.org/abs/1306.5541} {arXiv:1306.5541 [hep-ph]} \BibitemShut
  {NoStop}%
%%CITATION = ARXIV:1306.5541;%%
\bibitem [{\citenamefont {Han}\ \emph {et~al.}(2014)\citenamefont {Han},
  \citenamefont {Ji}, \citenamefont {Wu}, \citenamefont {Wu},\ and\
  \citenamefont {Yang}}]{Han:2013sga}%
  \BibitemOpen
  \bibfield  {author} {\bibinfo {author} {\bibfnamefont {C.}~\bibnamefont
  {Han}}, \bibinfo {author} {\bibfnamefont {X.}~\bibnamefont {Ji}}, \bibinfo
  {author} {\bibfnamefont {L.}~\bibnamefont {Wu}}, \bibinfo {author}
  {\bibfnamefont {P.}~\bibnamefont {Wu}}, \ and\ \bibinfo {author}
  {\bibfnamefont {J.~M.}\ \bibnamefont {Yang}},\ }\href {\doibase
  10.1007/JHEP04(2014)003} {\bibfield  {journal} {\bibinfo  {journal} {JHEP}\
  }\textbf {\bibinfo {volume} {1404}},\ \bibinfo {pages} {003} (\bibinfo {year}
  {2014})},\ \Eprint {http://arxiv.org/abs/1307.3790} {arXiv:1307.3790
  [hep-ph]} \BibitemShut {NoStop}%
%%CITATION = ARXIV:1307.3790;%%
\bibitem [{\citenamefont {No}\ and\ \citenamefont
  {Ramsey-Musolf}(2014)}]{No:2013wsa}%
  \BibitemOpen
  \bibfield  {author} {\bibinfo {author} {\bibfnamefont {J.~M.}\ \bibnamefont
  {No}}\ and\ \bibinfo {author} {\bibfnamefont {M.}~\bibnamefont
  {Ramsey-Musolf}},\ }\href {\doibase 10.1103/PhysRevD.89.095031} {\bibfield
  {journal} {\bibinfo  {journal} {Phys.Rev.}\ }\textbf {\bibinfo {volume}
  {D89}},\ \bibinfo {pages} {095031} (\bibinfo {year} {2014})},\ \Eprint
  {http://arxiv.org/abs/1310.6035} {arXiv:1310.6035 [hep-ph]} \BibitemShut
  {NoStop}%
%%CITATION = ARXIV:1310.6035;%%
\bibitem [{\citenamefont {McCullough}(2014)}]{McCullough:2013rea}%
  \BibitemOpen
  \bibfield  {author} {\bibinfo {author} {\bibfnamefont {M.}~\bibnamefont
  {McCullough}},\ }\href {\doibase 10.1103/PhysRevD.90.015001} {\bibfield
  {journal} {\bibinfo  {journal} {Phys.Rev.}\ }\textbf {\bibinfo {volume}
  {D90}},\ \bibinfo {pages} {015001} (\bibinfo {year} {2014})},\ \Eprint
  {http://arxiv.org/abs/1312.3322} {arXiv:1312.3322 [hep-ph]} \BibitemShut
  {NoStop}%
%%CITATION = ARXIV:1312.3322;%%
\bibitem [{\citenamefont {Grober}\ and\ \citenamefont
  {Muhlleitner}(2011)}]{Grober:2010yv}%
  \BibitemOpen
  \bibfield  {author} {\bibinfo {author} {\bibfnamefont {R.}~\bibnamefont
  {Grober}}\ and\ \bibinfo {author} {\bibfnamefont {M.}~\bibnamefont
  {Muhlleitner}},\ }\href {\doibase 10.1007/JHEP06(2011)020} {\bibfield
  {journal} {\bibinfo  {journal} {JHEP}\ }\textbf {\bibinfo {volume} {1106}},\
  \bibinfo {pages} {020} (\bibinfo {year} {2011})},\ \Eprint
  {http://arxiv.org/abs/1012.1562} {arXiv:1012.1562 [hep-ph]} \BibitemShut
  {NoStop}%
%%CITATION = ARXIV:1012.1562;%%
\bibitem [{\citenamefont {Contino}\ \emph {et~al.}(2012)\citenamefont
  {Contino}, \citenamefont {Ghezzi}, \citenamefont {Moretti}, \citenamefont
  {Panico}, \citenamefont {Piccinini} \emph {et~al.}}]{Contino:2012xk}%
  \BibitemOpen
  \bibfield  {author} {\bibinfo {author} {\bibfnamefont {R.}~\bibnamefont
  {Contino}}, \bibinfo {author} {\bibfnamefont {M.}~\bibnamefont {Ghezzi}},
  \bibinfo {author} {\bibfnamefont {M.}~\bibnamefont {Moretti}}, \bibinfo
  {author} {\bibfnamefont {G.}~\bibnamefont {Panico}}, \bibinfo {author}
  {\bibfnamefont {F.}~\bibnamefont {Piccinini}},  \emph {et~al.},\ }\href
  {\doibase 10.1007/JHEP08(2012)154} {\bibfield  {journal} {\bibinfo  {journal}
  {JHEP}\ }\textbf {\bibinfo {volume} {1208}},\ \bibinfo {pages} {154}
  (\bibinfo {year} {2012})},\ \Eprint {http://arxiv.org/abs/1205.5444}
  {arXiv:1205.5444 [hep-ph]} \BibitemShut {NoStop}%
%%CITATION = ARXIV:1205.5444;%%
\bibitem [{\citenamefont {Gillioz}\ \emph {et~al.}(2012)\citenamefont
  {Gillioz}, \citenamefont {Grober}, \citenamefont {Grojean}, \citenamefont
  {Muhlleitner},\ and\ \citenamefont {Salvioni}}]{Gillioz:2012se}%
  \BibitemOpen
  \bibfield  {author} {\bibinfo {author} {\bibfnamefont {M.}~\bibnamefont
  {Gillioz}}, \bibinfo {author} {\bibfnamefont {R.}~\bibnamefont {Grober}},
  \bibinfo {author} {\bibfnamefont {C.}~\bibnamefont {Grojean}}, \bibinfo
  {author} {\bibfnamefont {M.}~\bibnamefont {Muhlleitner}}, \ and\ \bibinfo
  {author} {\bibfnamefont {E.}~\bibnamefont {Salvioni}},\ }\href {\doibase
  10.1007/JHEP10(2012)004} {\bibfield  {journal} {\bibinfo  {journal} {JHEP}\
  }\textbf {\bibinfo {volume} {1210}},\ \bibinfo {pages} {004} (\bibinfo {year}
  {2012})},\ \Eprint {http://arxiv.org/abs/1206.7120} {arXiv:1206.7120
  [hep-ph]} \BibitemShut {NoStop}%
%%CITATION = ARXIV:1206.7120;%%
\bibitem [{\citenamefont {Kribs}\ and\ \citenamefont
  {Martin}(2012)}]{Kribs:2012kz}%
  \BibitemOpen
  \bibfield  {author} {\bibinfo {author} {\bibfnamefont {G.~D.}\ \bibnamefont
  {Kribs}}\ and\ \bibinfo {author} {\bibfnamefont {A.}~\bibnamefont {Martin}},\
  }\href {\doibase 10.1103/PhysRevD.86.095023} {\bibfield  {journal} {\bibinfo
  {journal} {Phys.Rev.}\ }\textbf {\bibinfo {volume} {D86}},\ \bibinfo {pages}
  {095023} (\bibinfo {year} {2012})},\ \Eprint {http://arxiv.org/abs/1207.4496}
  {arXiv:1207.4496 [hep-ph]} \BibitemShut {NoStop}%
%%CITATION = ARXIV:1207.4496;%%
\bibitem [{\citenamefont {Dawson}\ \emph {et~al.}(2013)\citenamefont {Dawson},
  \citenamefont {Furlan},\ and\ \citenamefont {Lewis}}]{Dawson:2012mk}%
  \BibitemOpen
  \bibfield  {author} {\bibinfo {author} {\bibfnamefont {S.}~\bibnamefont
  {Dawson}}, \bibinfo {author} {\bibfnamefont {E.}~\bibnamefont {Furlan}}, \
  and\ \bibinfo {author} {\bibfnamefont {I.}~\bibnamefont {Lewis}},\ }\href
  {\doibase 10.1103/PhysRevD.87.014007} {\bibfield  {journal} {\bibinfo
  {journal} {Phys.Rev.}\ }\textbf {\bibinfo {volume} {D87}},\ \bibinfo {pages}
  {014007} (\bibinfo {year} {2013})},\ \Eprint {http://arxiv.org/abs/1210.6663}
  {arXiv:1210.6663 [hep-ph]} \BibitemShut {NoStop}%
%%CITATION = ARXIV:1210.6663;%%
\bibitem [{\citenamefont {Chen}\ \emph {et~al.}(2014)\citenamefont {Chen},
  \citenamefont {Dawson},\ and\ \citenamefont {Lewis}}]{Chen:2014xwa}%
  \BibitemOpen
  \bibfield  {author} {\bibinfo {author} {\bibfnamefont {C.-Y.}\ \bibnamefont
  {Chen}}, \bibinfo {author} {\bibfnamefont {S.}~\bibnamefont {Dawson}}, \ and\
  \bibinfo {author} {\bibfnamefont {I.}~\bibnamefont {Lewis}},\ }\href
  {\doibase 10.1103/PhysRevD.90.035016} {\bibfield  {journal} {\bibinfo
  {journal} {Phys.Rev.}\ }\textbf {\bibinfo {volume} {D90}},\ \bibinfo {pages}
  {035016} (\bibinfo {year} {2014})},\ \Eprint {http://arxiv.org/abs/1406.3349}
  {arXiv:1406.3349 [hep-ph]} \BibitemShut {NoStop}%
%%CITATION = ARXIV:1406.3349;%%
\bibitem [{\citenamefont {Nishiwaki}\ \emph {et~al.}(2014)\citenamefont
  {Nishiwaki}, \citenamefont {Niyogi},\ and\ \citenamefont
  {Shivaji}}]{Nishiwaki:2013cma}%
  \BibitemOpen
  \bibfield  {author} {\bibinfo {author} {\bibfnamefont {K.}~\bibnamefont
  {Nishiwaki}}, \bibinfo {author} {\bibfnamefont {S.}~\bibnamefont {Niyogi}}, \
  and\ \bibinfo {author} {\bibfnamefont {A.}~\bibnamefont {Shivaji}},\ }\href
  {\doibase 10.1007/JHEP04(2014)011} {\bibfield  {journal} {\bibinfo  {journal}
  {JHEP}\ }\textbf {\bibinfo {volume} {1404}},\ \bibinfo {pages} {011}
  (\bibinfo {year} {2014})},\ \Eprint {http://arxiv.org/abs/1309.6907}
  {arXiv:1309.6907 [hep-ph]} \BibitemShut {NoStop}%
%%CITATION = ARXIV:1309.6907;%%
\bibitem [{\citenamefont {Liu}\ \emph {et~al.}(2013)\citenamefont {Liu},
  \citenamefont {Wang},\ and\ \citenamefont {Zhu}}]{Liu:2013woa}%
  \BibitemOpen
  \bibfield  {author} {\bibinfo {author} {\bibfnamefont {J.}~\bibnamefont
  {Liu}}, \bibinfo {author} {\bibfnamefont {X.-P.}\ \bibnamefont {Wang}}, \
  and\ \bibinfo {author} {\bibfnamefont {S.-h.}\ \bibnamefont {Zhu}},\
  }\href@noop {} {\  (\bibinfo {year} {2013})},\ \Eprint
  {http://arxiv.org/abs/1310.3634} {arXiv:1310.3634 [hep-ph]} \BibitemShut
  {NoStop}%
%%CITATION = ARXIV:1310.3634;%%
\bibitem [{\citenamefont {Enkhbat}(2014)}]{Enkhbat:2013oba}%
  \BibitemOpen
  \bibfield  {author} {\bibinfo {author} {\bibfnamefont {T.}~\bibnamefont
  {Enkhbat}},\ }\href {\doibase 10.1007/JHEP01(2014)158} {\bibfield  {journal}
  {\bibinfo  {journal} {JHEP}\ }\textbf {\bibinfo {volume} {1401}},\ \bibinfo
  {pages} {158} (\bibinfo {year} {2014})},\ \Eprint
  {http://arxiv.org/abs/1311.4445} {arXiv:1311.4445 [hep-ph]} \BibitemShut
  {NoStop}%
%%CITATION = ARXIV:1311.4445;%%
\bibitem [{\citenamefont {Heng}\ \emph {et~al.}(2014)\citenamefont {Heng},
  \citenamefont {Shang}, \citenamefont {Zhang},\ and\ \citenamefont
  {Zhu}}]{Heng:2013cya}%
  \BibitemOpen
  \bibfield  {author} {\bibinfo {author} {\bibfnamefont {Z.}~\bibnamefont
  {Heng}}, \bibinfo {author} {\bibfnamefont {L.}~\bibnamefont {Shang}},
  \bibinfo {author} {\bibfnamefont {Y.}~\bibnamefont {Zhang}}, \ and\ \bibinfo
  {author} {\bibfnamefont {J.}~\bibnamefont {Zhu}},\ }\href {\doibase
  10.1007/JHEP02(2014)083} {\bibfield  {journal} {\bibinfo  {journal} {JHEP}\
  }\textbf {\bibinfo {volume} {1402}},\ \bibinfo {pages} {083} (\bibinfo {year}
  {2014})},\ \Eprint {http://arxiv.org/abs/1312.4260} {arXiv:1312.4260
  [hep-ph]} \BibitemShut {NoStop}%
%%CITATION = ARXIV:1312.4260;%%
\bibitem [{\citenamefont {Frederix}\ \emph {et~al.}(2014)\citenamefont
  {Frederix}, \citenamefont {Frixione}, \citenamefont {Hirschi}, \citenamefont
  {Maltoni}, \citenamefont {Mattelaer} \emph {et~al.}}]{Frederix:2014hta}%
  \BibitemOpen
  \bibfield  {author} {\bibinfo {author} {\bibfnamefont {R.}~\bibnamefont
  {Frederix}}, \bibinfo {author} {\bibfnamefont {S.}~\bibnamefont {Frixione}},
  \bibinfo {author} {\bibfnamefont {V.}~\bibnamefont {Hirschi}}, \bibinfo
  {author} {\bibfnamefont {F.}~\bibnamefont {Maltoni}}, \bibinfo {author}
  {\bibfnamefont {O.}~\bibnamefont {Mattelaer}},  \emph {et~al.},\ }\href
  {\doibase 10.1016/j.physletb.2014.03.026} {\bibfield  {journal} {\bibinfo
  {journal} {Phys.Lett.}\ }\textbf {\bibinfo {volume} {B732}},\ \bibinfo
  {pages} {142} (\bibinfo {year} {2014})},\ \Eprint
  {http://arxiv.org/abs/1401.7340} {arXiv:1401.7340 [hep-ph]} \BibitemShut
  {NoStop}%
%%CITATION = ARXIV:1401.7340;%%
\bibitem [{\citenamefont {Baglio}\ \emph {et~al.}(2014)\citenamefont {Baglio},
  \citenamefont {Eberhardt}, \citenamefont {Nierste},\ and\ \citenamefont
  {Wiebusch}}]{Baglio:2014nea}%
  \BibitemOpen
  \bibfield  {author} {\bibinfo {author} {\bibfnamefont {J.}~\bibnamefont
  {Baglio}}, \bibinfo {author} {\bibfnamefont {O.}~\bibnamefont {Eberhardt}},
  \bibinfo {author} {\bibfnamefont {U.}~\bibnamefont {Nierste}}, \ and\
  \bibinfo {author} {\bibfnamefont {M.}~\bibnamefont {Wiebusch}},\ }\href
  {\doibase 10.1103/PhysRevD.90.015008} {\bibfield  {journal} {\bibinfo
  {journal} {Phys.Rev.}\ }\textbf {\bibinfo {volume} {D90}},\ \bibinfo {pages}
  {015008} (\bibinfo {year} {2014})},\ \Eprint {http://arxiv.org/abs/1403.1264}
  {arXiv:1403.1264 [hep-ph]} \BibitemShut {NoStop}%
%%CITATION = ARXIV:1403.1264;%%
\bibitem [{\citenamefont {Hespel}\ \emph {et~al.}(2014)\citenamefont {Hespel},
  \citenamefont {Lopez-Val},\ and\ \citenamefont {Vryonidou}}]{Hespel:2014sla}%
  \BibitemOpen
  \bibfield  {author} {\bibinfo {author} {\bibfnamefont {B.}~\bibnamefont
  {Hespel}}, \bibinfo {author} {\bibfnamefont {D.}~\bibnamefont {Lopez-Val}}, \
  and\ \bibinfo {author} {\bibfnamefont {E.}~\bibnamefont {Vryonidou}},\ }\href
  {\doibase 10.1007/JHEP09(2014)124} {\bibfield  {journal} {\bibinfo  {journal}
  {JHEP}\ }\textbf {\bibinfo {volume} {1409}},\ \bibinfo {pages} {124}
  (\bibinfo {year} {2014})},\ \Eprint {http://arxiv.org/abs/1407.0281}
  {arXiv:1407.0281 [hep-ph]} \BibitemShut {NoStop}%
%%CITATION = ARXIV:1407.0281;%%
\bibitem [{\citenamefont {Bhattacherjee}\ and\ \citenamefont
  {Choudhury}(2014)}]{Bhattacherjee:2014bca}%
  \BibitemOpen
  \bibfield  {author} {\bibinfo {author} {\bibfnamefont {B.}~\bibnamefont
  {Bhattacherjee}}\ and\ \bibinfo {author} {\bibfnamefont {A.}~\bibnamefont
  {Choudhury}},\ }\href@noop {} {\  (\bibinfo {year} {2014})},\ \Eprint
  {http://arxiv.org/abs/1407.6866} {arXiv:1407.6866 [hep-ph]} \BibitemShut
  {NoStop}%
%%CITATION = ARXIV:1407.6866;%%
\bibitem [{\citenamefont {Liu}\ \emph {et~al.}(2015)\citenamefont {Liu},
  \citenamefont {Hu}, \citenamefont {Yang},\ and\ \citenamefont
  {Han}}]{Liu:2014rba}%
  \BibitemOpen
  \bibfield  {author} {\bibinfo {author} {\bibfnamefont {N.}~\bibnamefont
  {Liu}}, \bibinfo {author} {\bibfnamefont {S.}~\bibnamefont {Hu}}, \bibinfo
  {author} {\bibfnamefont {B.}~\bibnamefont {Yang}}, \ and\ \bibinfo {author}
  {\bibfnamefont {J.}~\bibnamefont {Han}},\ }\href {\doibase
  10.1007/JHEP01(2015)008} {\bibfield  {journal} {\bibinfo  {journal} {JHEP}\
  }\textbf {\bibinfo {volume} {1501}},\ \bibinfo {pages} {008} (\bibinfo {year}
  {2015})},\ \Eprint {http://arxiv.org/abs/1408.4191} {arXiv:1408.4191
  [hep-ph]} \BibitemShut {NoStop}%
%%CITATION = ARXIV:1408.4191;%%
\bibitem [{\citenamefont {Cao}\ \emph {et~al.}(2014)\citenamefont {Cao},
  \citenamefont {Li}, \citenamefont {Shang}, \citenamefont {Wu},\ and\
  \citenamefont {Zhang}}]{Cao:2014kya}%
  \BibitemOpen
  \bibfield  {author} {\bibinfo {author} {\bibfnamefont {J.}~\bibnamefont
  {Cao}}, \bibinfo {author} {\bibfnamefont {D.}~\bibnamefont {Li}}, \bibinfo
  {author} {\bibfnamefont {L.}~\bibnamefont {Shang}}, \bibinfo {author}
  {\bibfnamefont {P.}~\bibnamefont {Wu}}, \ and\ \bibinfo {author}
  {\bibfnamefont {Y.}~\bibnamefont {Zhang}},\ }\href {\doibase
  10.1007/JHEP12(2014)026} {\bibfield  {journal} {\bibinfo  {journal} {JHEP}\
  }\textbf {\bibinfo {volume} {1412}},\ \bibinfo {pages} {026} (\bibinfo {year}
  {2014})},\ \Eprint {http://arxiv.org/abs/1409.8431} {arXiv:1409.8431
  [hep-ph]} \BibitemShut {NoStop}%
%%CITATION = ARXIV:1409.8431;%%
\bibitem [{\citenamefont {Maltoni}\ \emph {et~al.}(2014)\citenamefont
  {Maltoni}, \citenamefont {Vryonidou},\ and\ \citenamefont
  {Zaro}}]{Maltoni:2014eza}%
  \BibitemOpen
  \bibfield  {author} {\bibinfo {author} {\bibfnamefont {F.}~\bibnamefont
  {Maltoni}}, \bibinfo {author} {\bibfnamefont {E.}~\bibnamefont {Vryonidou}},
  \ and\ \bibinfo {author} {\bibfnamefont {M.}~\bibnamefont {Zaro}},\ }\href
  {\doibase 10.1007/JHEP11(2014)079} {\bibfield  {journal} {\bibinfo  {journal}
  {JHEP}\ }\textbf {\bibinfo {volume} {1411}},\ \bibinfo {pages} {079}
  (\bibinfo {year} {2014})},\ \Eprint {http://arxiv.org/abs/1408.6542}
  {arXiv:1408.6542 [hep-ph]} \BibitemShut {NoStop}%
%%CITATION = ARXIV:1408.6542;%%
\bibitem [{\citenamefont {Martin-Lozano}\ \emph {et~al.}(2015)\citenamefont
  {Martin-Lozano}, \citenamefont {Moreno},\ and\ \citenamefont
  {Park}}]{Martin-Lozano:2015dja}%
  \BibitemOpen
  \bibfield  {author} {\bibinfo {author} {\bibfnamefont {V.}~\bibnamefont
  {Martin-Lozano}}, \bibinfo {author} {\bibfnamefont {J.~M.}\ \bibnamefont
  {Moreno}}, \ and\ \bibinfo {author} {\bibfnamefont {C.~B.}\ \bibnamefont
  {Park}},\ }\href@noop {} {\  (\bibinfo {year} {2015})},\ \Eprint
  {http://arxiv.org/abs/1501.03799} {arXiv:1501.03799 [hep-ph]} \BibitemShut
  {NoStop}%
%%CITATION = ARXIV:1501.03799;%%
\bibitem [{\citenamefont {van Beekveld}\ \emph {et~al.}(2015)\citenamefont {van
  Beekveld}, \citenamefont {Beenakker}, \citenamefont {Caron}, \citenamefont
  {Castelijn}, \citenamefont {Lanfermann} \emph
  {et~al.}}]{vanBeekveld:2015tka}%
  \BibitemOpen
  \bibfield  {author} {\bibinfo {author} {\bibfnamefont {M.}~\bibnamefont {van
  Beekveld}}, \bibinfo {author} {\bibfnamefont {W.}~\bibnamefont {Beenakker}},
  \bibinfo {author} {\bibfnamefont {S.}~\bibnamefont {Caron}}, \bibinfo
  {author} {\bibfnamefont {R.}~\bibnamefont {Castelijn}}, \bibinfo {author}
  {\bibfnamefont {M.}~\bibnamefont {Lanfermann}},  \emph {et~al.},\ }\href
  {\doibase 10.1007/JHEP05(2015)044} {\bibfield  {journal} {\bibinfo  {journal}
  {JHEP}\ }\textbf {\bibinfo {volume} {1505}},\ \bibinfo {pages} {044}
  (\bibinfo {year} {2015})},\ \Eprint {http://arxiv.org/abs/1501.02145v2}
  {arXiv:1501.02145v2 [hep-ph]} \BibitemShut {NoStop}%
%%CITATION = ARXIV:1501.02145V2;%%
\bibitem [{\citenamefont {Goertz}\ \emph {et~al.}(2014)\citenamefont {Goertz},
  \citenamefont {Papaefstathiou}, \citenamefont {Yang},\ and\ \citenamefont
  {Zurita}}]{Goertz:2014qta}%
  \BibitemOpen
  \bibfield  {author} {\bibinfo {author} {\bibfnamefont {F.}~\bibnamefont
  {Goertz}}, \bibinfo {author} {\bibfnamefont {A.}~\bibnamefont
  {Papaefstathiou}}, \bibinfo {author} {\bibfnamefont {L.~L.}\ \bibnamefont
  {Yang}}, \ and\ \bibinfo {author} {\bibfnamefont {J.}~\bibnamefont
  {Zurita}},\ }\href@noop {} {\  (\bibinfo {year} {2014})},\ \Eprint
  {http://arxiv.org/abs/1410.3471} {arXiv:1410.3471 [hep-ph]} \BibitemShut
  {NoStop}%
%%CITATION = ARXIV:1410.3471;%%
\bibitem [{\citenamefont {Azatov}\ \emph {et~al.}(2015)\citenamefont {Azatov},
  \citenamefont {Contino}, \citenamefont {Panico},\ and\ \citenamefont
  {Son}}]{Azatov:2015oxa}%
  \BibitemOpen
  \bibfield  {author} {\bibinfo {author} {\bibfnamefont {A.}~\bibnamefont
  {Azatov}}, \bibinfo {author} {\bibfnamefont {R.}~\bibnamefont {Contino}},
  \bibinfo {author} {\bibfnamefont {G.}~\bibnamefont {Panico}}, \ and\ \bibinfo
  {author} {\bibfnamefont {M.}~\bibnamefont {Son}},\ }\href@noop {} {\
  (\bibinfo {year} {2015})},\ \Eprint {http://arxiv.org/abs/1502.00539}
  {arXiv:1502.00539 [hep-ph]} \BibitemShut {NoStop}%
%%CITATION = ARXIV:1502.00539;%%
\bibitem [{\citenamefont {Goertz}(2015)}]{Goertz:2015dba}%
  \BibitemOpen
  \bibfield  {author} {\bibinfo {author} {\bibfnamefont {F.}~\bibnamefont
  {Goertz}},\ }\href@noop {} {\  (\bibinfo {year} {2015})},\ \Eprint
  {http://arxiv.org/abs/1504.00355} {arXiv:1504.00355 [hep-ph]} \BibitemShut
  {NoStop}%
%%CITATION = ARXIV:1504.00355;%%
\bibitem [{\citenamefont {Aad}\ \emph {et~al.}(2015)\citenamefont {Aad} \emph
  {et~al.}}]{Aad:2014yja}%
  \BibitemOpen
  \bibfield  {author} {\bibinfo {author} {\bibfnamefont {G.}~\bibnamefont
  {Aad}} \emph {et~al.} (\bibinfo {collaboration} {ATLAS}),\ }\href {\doibase
  10.1103/PhysRevLett.114.081802} {\bibfield  {journal} {\bibinfo  {journal}
  {Phys.Rev.Lett.}\ }\textbf {\bibinfo {volume} {114}},\ \bibinfo {pages}
  {081802} (\bibinfo {year} {2015})},\ \Eprint {http://arxiv.org/abs/1406.5053}
  {arXiv:1406.5053 [hep-ex]} \BibitemShut {NoStop}%
%%CITATION = ARXIV:1406.5053;%%
\bibitem [{CMS(2014)}]{CMS-PAS-HIG-13-032}%
  \BibitemOpen
  \href {http://cds.cern.ch/record/1697512} {\emph {\bibinfo {title} {{Search
  for resonant HH production in 2gamma+2b channel}}}},\ \bibinfo {type} {Tech.
  Rep.}\ \bibinfo {number} {CMS-PAS-HIG-13-032}\ (\bibinfo  {institution}
  {CERN},\ \bibinfo {address} {Geneva},\ \bibinfo {year} {2014})\BibitemShut
  {NoStop}%
\bibitem [{\citenamefont {Khachatryan}\ \emph {et~al.}(2015)\citenamefont
  {Khachatryan} \emph {et~al.}}]{Khachatryan:2015yea}%
  \BibitemOpen
  \bibfield  {author} {\bibinfo {author} {\bibfnamefont {V.}~\bibnamefont
  {Khachatryan}} \emph {et~al.} (\bibinfo {collaboration} {CMS}),\ }\href@noop
  {} {\  (\bibinfo {year} {2015})},\ \Eprint {http://arxiv.org/abs/1503.04114}
  {arXiv:1503.04114 [hep-ex]} \BibitemShut {NoStop}%
%%CITATION = ARXIV:1503.04114;%%
\bibitem [{\citenamefont {Barr}\ \emph {et~al.}(2015)\citenamefont {Barr},
  \citenamefont {Dolan}, \citenamefont {Englert}, \citenamefont {Ferreira~de
  Lima},\ and\ \citenamefont {Spannowsky}}]{Barr:2014sga}%
  \BibitemOpen
  \bibfield  {author} {\bibinfo {author} {\bibfnamefont {A.~J.}\ \bibnamefont
  {Barr}}, \bibinfo {author} {\bibfnamefont {M.~J.}\ \bibnamefont {Dolan}},
  \bibinfo {author} {\bibfnamefont {C.}~\bibnamefont {Englert}}, \bibinfo
  {author} {\bibfnamefont {D.~E.}\ \bibnamefont {Ferreira~de Lima}}, \ and\
  \bibinfo {author} {\bibfnamefont {M.}~\bibnamefont {Spannowsky}},\ }\href
  {\doibase 10.1007/JHEP02(2015)016} {\bibfield  {journal} {\bibinfo  {journal}
  {JHEP}\ }\textbf {\bibinfo {volume} {1502}},\ \bibinfo {pages} {016}
  (\bibinfo {year} {2015})},\ \Eprint {http://arxiv.org/abs/1412.7154}
  {arXiv:1412.7154 [hep-ph]} \BibitemShut {NoStop}%
%%CITATION = ARXIV:1412.7154;%%
\bibitem [{\citenamefont {Li}\ \emph {et~al.}(2015)\citenamefont {Li},
  \citenamefont {Li}, \citenamefont {Yan},\ and\ \citenamefont
  {Zhao}}]{Li:2015yia}%
  \BibitemOpen
  \bibfield  {author} {\bibinfo {author} {\bibfnamefont {Q.}~\bibnamefont
  {Li}}, \bibinfo {author} {\bibfnamefont {Z.}~\bibnamefont {Li}}, \bibinfo
  {author} {\bibfnamefont {Q.-S.}\ \bibnamefont {Yan}}, \ and\ \bibinfo
  {author} {\bibfnamefont {X.}~\bibnamefont {Zhao}},\ }\href@noop {} {\
  (\bibinfo {year} {2015})},\ \Eprint {http://arxiv.org/abs/1503.07611}
  {arXiv:1503.07611 [hep-ph]} \BibitemShut {NoStop}%
%%CITATION = ARXIV:1503.07611;%%
\bibitem [{\citenamefont {de~Florian}\ and\ \citenamefont
  {Mazzitelli}(2013{\natexlab{a}})}]{deFlorian:2013jea}%
  \BibitemOpen
  \bibfield  {author} {\bibinfo {author} {\bibfnamefont {D.}~\bibnamefont
  {de~Florian}}\ and\ \bibinfo {author} {\bibfnamefont {J.}~\bibnamefont
  {Mazzitelli}},\ }\href@noop {} {\  (\bibinfo {year} {2013}{\natexlab{a}})},\
  \Eprint {http://arxiv.org/abs/1309.6594} {arXiv:1309.6594 [hep-ph]}
  \BibitemShut {NoStop}%
%%CITATION = ARXIV:1309.6594;%%
\bibitem [{\citenamefont {Glover}\ and\ \citenamefont {van~der
  Bij}(1988)}]{Glover:1987nx}%
  \BibitemOpen
  \bibfield  {author} {\bibinfo {author} {\bibfnamefont {E.~N.}\ \bibnamefont
  {Glover}}\ and\ \bibinfo {author} {\bibfnamefont {J.}~\bibnamefont {van~der
  Bij}},\ }\href {\doibase 10.1016/0550-3213(88)90083-1} {\bibfield  {journal}
  {\bibinfo  {journal} {Nucl.Phys.}\ }\textbf {\bibinfo {volume} {B309}},\
  \bibinfo {pages} {282} (\bibinfo {year} {1988})}\BibitemShut {NoStop}%
%%CITATION = NUPHA,B309,282;%%
\bibitem [{\citenamefont {Dawson}\ \emph {et~al.}(1998)\citenamefont {Dawson},
  \citenamefont {Dittmaier},\ and\ \citenamefont {Spira}}]{Dawson:1998py}%
  \BibitemOpen
  \bibfield  {author} {\bibinfo {author} {\bibfnamefont {S.}~\bibnamefont
  {Dawson}}, \bibinfo {author} {\bibfnamefont {S.}~\bibnamefont {Dittmaier}}, \
  and\ \bibinfo {author} {\bibfnamefont {M.}~\bibnamefont {Spira}},\ }\href
  {\doibase 10.1103/PhysRevD.58.115012} {\bibfield  {journal} {\bibinfo
  {journal} {Phys.Rev.}\ }\textbf {\bibinfo {volume} {D58}},\ \bibinfo {pages}
  {115012} (\bibinfo {year} {1998})},\ \Eprint
  {http://arxiv.org/abs/hep-ph/9805244} {arXiv:hep-ph/9805244 [hep-ph]}
  \BibitemShut {NoStop}%
%%CITATION = HEP-PH/9805244;%%
\bibitem [{\citenamefont {Djouadi}\ \emph {et~al.}(1999)\citenamefont
  {Djouadi}, \citenamefont {Kilian}, \citenamefont {Muhlleitner},\ and\
  \citenamefont {Zerwas}}]{Djouadi:1999rca}%
  \BibitemOpen
  \bibfield  {author} {\bibinfo {author} {\bibfnamefont {A.}~\bibnamefont
  {Djouadi}}, \bibinfo {author} {\bibfnamefont {W.}~\bibnamefont {Kilian}},
  \bibinfo {author} {\bibfnamefont {M.}~\bibnamefont {Muhlleitner}}, \ and\
  \bibinfo {author} {\bibfnamefont {P.}~\bibnamefont {Zerwas}},\ }\href
  {\doibase 10.1007/s100529900083} {\bibfield  {journal} {\bibinfo  {journal}
  {Eur.Phys.J.}\ }\textbf {\bibinfo {volume} {C10}},\ \bibinfo {pages} {45}
  (\bibinfo {year} {1999})},\ \Eprint {http://arxiv.org/abs/hep-ph/9904287}
  {arXiv:hep-ph/9904287 [hep-ph]} \BibitemShut {NoStop}%
%%CITATION = HEP-PH/9904287;%%
\bibitem [{\citenamefont {Plehn}\ \emph {et~al.}(1996)\citenamefont {Plehn},
  \citenamefont {Spira},\ and\ \citenamefont {Zerwas}}]{Plehn:1996wb}%
  \BibitemOpen
  \bibfield  {author} {\bibinfo {author} {\bibfnamefont {T.}~\bibnamefont
  {Plehn}}, \bibinfo {author} {\bibfnamefont {M.}~\bibnamefont {Spira}}, \ and\
  \bibinfo {author} {\bibfnamefont {P.}~\bibnamefont {Zerwas}},\ }\href
  {\doibase 10.1016/0550-3213(96)00418-X} {\bibfield  {journal} {\bibinfo
  {journal} {Nucl.Phys.}\ }\textbf {\bibinfo {volume} {B479}},\ \bibinfo
  {pages} {46} (\bibinfo {year} {1996})},\ \Eprint
  {http://arxiv.org/abs/hep-ph/9603205} {arXiv:hep-ph/9603205 [hep-ph]}
  \BibitemShut {NoStop}%
%%CITATION = HEP-PH/9603205;%%
\bibitem [{\citenamefont {de~Florian}\ and\ \citenamefont
  {Mazzitelli}(2013{\natexlab{b}})}]{deFlorian:2013uza}%
  \BibitemOpen
  \bibfield  {author} {\bibinfo {author} {\bibfnamefont {D.}~\bibnamefont
  {de~Florian}}\ and\ \bibinfo {author} {\bibfnamefont {J.}~\bibnamefont
  {Mazzitelli}},\ }\href {\doibase 10.1016/j.physletb.2013.06.046} {\bibfield
  {journal} {\bibinfo  {journal} {Phys.Lett.}\ }\textbf {\bibinfo {volume}
  {B724}},\ \bibinfo {pages} {306} (\bibinfo {year} {2013}{\natexlab{b}})},\
  \Eprint {http://arxiv.org/abs/1305.5206} {arXiv:1305.5206 [hep-ph]}
  \BibitemShut {NoStop}%
%%CITATION = ARXIV:1305.5206;%%
\bibitem [{\citenamefont {Grigo}\ \emph {et~al.}(2013)\citenamefont {Grigo},
  \citenamefont {Hoff}, \citenamefont {Melnikov},\ and\ \citenamefont
  {Steinhauser}}]{Grigo:2013rya}%
  \BibitemOpen
  \bibfield  {author} {\bibinfo {author} {\bibfnamefont {J.}~\bibnamefont
  {Grigo}}, \bibinfo {author} {\bibfnamefont {J.}~\bibnamefont {Hoff}},
  \bibinfo {author} {\bibfnamefont {K.}~\bibnamefont {Melnikov}}, \ and\
  \bibinfo {author} {\bibfnamefont {M.}~\bibnamefont {Steinhauser}},\ }\href
  {\doibase 10.1016/j.nuclphysb.2013.06.024} {\bibfield  {journal} {\bibinfo
  {journal} {Nucl.Phys.}\ }\textbf {\bibinfo {volume} {B875}},\ \bibinfo
  {pages} {1} (\bibinfo {year} {2013})},\ \Eprint
  {http://arxiv.org/abs/1305.7340} {arXiv:1305.7340 [hep-ph]} \BibitemShut
  {NoStop}%
%%CITATION = ARXIV:1305.7340;%%
\bibitem [{\citenamefont {de~Florian}\ and\ \citenamefont
  {Mazzitelli}(2015)}]{deFlorian:2015moa}%
  \BibitemOpen
  \bibfield  {author} {\bibinfo {author} {\bibfnamefont {D.}~\bibnamefont
  {de~Florian}}\ and\ \bibinfo {author} {\bibfnamefont {J.}~\bibnamefont
  {Mazzitelli}},\ }\href@noop {} {\  (\bibinfo {year} {2015})},\ \Eprint
  {http://arxiv.org/abs/1505.07122} {arXiv:1505.07122 [hep-ph]} \BibitemShut
  {NoStop}%
%%CITATION = ARXIV:1505.07122;%%
\bibitem [{\citenamefont {Actis}\ \emph {et~al.}(2008)\citenamefont {Actis},
  \citenamefont {Passarino}, \citenamefont {Sturm},\ and\ \citenamefont
  {Uccirati}}]{Actis:2008ug}%
  \BibitemOpen
  \bibfield  {author} {\bibinfo {author} {\bibfnamefont {S.}~\bibnamefont
  {Actis}}, \bibinfo {author} {\bibfnamefont {G.}~\bibnamefont {Passarino}},
  \bibinfo {author} {\bibfnamefont {C.}~\bibnamefont {Sturm}}, \ and\ \bibinfo
  {author} {\bibfnamefont {S.}~\bibnamefont {Uccirati}},\ }\href {\doibase
  10.1016/j.physletb.2008.10.018} {\bibfield  {journal} {\bibinfo  {journal}
  {Phys.Lett.}\ }\textbf {\bibinfo {volume} {B670}},\ \bibinfo {pages} {12}
  (\bibinfo {year} {2008})},\ \Eprint {http://arxiv.org/abs/0809.1301}
  {arXiv:0809.1301 [hep-ph]} \BibitemShut {NoStop}%
%%CITATION = ARXIV:0809.1301;%%
\bibitem [{The(2013{\natexlab{a}})}]{TheATLAScollaboration:performance2}%
  \BibitemOpen
  \href@noop {} {\bibfield  {journal} {\bibinfo  {journal}
  {ATL-PHYS-PUB-2013-009}\ } (\bibinfo {year}
  {2013}{\natexlab{a}})}\BibitemShut {NoStop}%
%%CITATION = ATL-PHYS-PUB-2013-009;%%
\bibitem [{The(2013{\natexlab{b}})}]{TheATLAScollaboration:performance1}%
  \BibitemOpen
  \href@noop {} {\bibfield  {journal} {\bibinfo  {journal}
  {ATL-PHYS-PUB-2013-004}\ } (\bibinfo {year}
  {2013}{\natexlab{b}})}\BibitemShut {NoStop}%
%%CITATION = ATL-PHYS-PUB-2013-004;%%
\bibitem [{\citenamefont {Cacciari}\ \emph {et~al.}(2012)\citenamefont
  {Cacciari}, \citenamefont {Salam},\ and\ \citenamefont
  {Soyez}}]{Cacciari:2011ma}%
  \BibitemOpen
  \bibfield  {author} {\bibinfo {author} {\bibfnamefont {M.}~\bibnamefont
  {Cacciari}}, \bibinfo {author} {\bibfnamefont {G.~P.}\ \bibnamefont {Salam}},
  \ and\ \bibinfo {author} {\bibfnamefont {G.}~\bibnamefont {Soyez}},\ }\href
  {\doibase 10.1140/epjc/s10052-012-1896-2} {\bibfield  {journal} {\bibinfo
  {journal} {Eur.Phys.J.}\ }\textbf {\bibinfo {volume} {C72}},\ \bibinfo
  {pages} {1896} (\bibinfo {year} {2012})},\ \Eprint
  {http://arxiv.org/abs/1111.6097} {arXiv:1111.6097 [hep-ph]} \BibitemShut
  {NoStop}%
%%CITATION = ARXIV:1111.6097;%%
\bibitem [{\citenamefont {Cacciari}\ and\ \citenamefont
  {Salam}(2006)}]{Cacciari:2005hq}%
  \BibitemOpen
  \bibfield  {author} {\bibinfo {author} {\bibfnamefont {M.}~\bibnamefont
  {Cacciari}}\ and\ \bibinfo {author} {\bibfnamefont {G.~P.}\ \bibnamefont
  {Salam}},\ }\href {\doibase 10.1016/j.physletb.2006.08.037} {\bibfield
  {journal} {\bibinfo  {journal} {Phys.Lett.}\ }\textbf {\bibinfo {volume}
  {B641}},\ \bibinfo {pages} {57} (\bibinfo {year} {2006})},\ \Eprint
  {http://arxiv.org/abs/hep-ph/0512210} {arXiv:hep-ph/0512210 [hep-ph]}
  \BibitemShut {NoStop}%
%%CITATION = HEP-PH/0512210;%%
\bibitem [{\citenamefont {Bahr}\ \emph {et~al.}(2008)\citenamefont {Bahr},
  \citenamefont {Gieseke}, \citenamefont {Gigg}, \citenamefont {Grellscheid},
  \citenamefont {Hamilton} \emph {et~al.}}]{Bahr:2008pv}%
  \BibitemOpen
  \bibfield  {author} {\bibinfo {author} {\bibfnamefont {M.}~\bibnamefont
  {Bahr}}, \bibinfo {author} {\bibfnamefont {S.}~\bibnamefont {Gieseke}},
  \bibinfo {author} {\bibfnamefont {M.}~\bibnamefont {Gigg}}, \bibinfo {author}
  {\bibfnamefont {D.}~\bibnamefont {Grellscheid}}, \bibinfo {author}
  {\bibfnamefont {K.}~\bibnamefont {Hamilton}},  \emph {et~al.},\ }\href
  {\doibase 10.1140/epjc/s10052-008-0798-9} {\bibfield  {journal} {\bibinfo
  {journal} {Eur.Phys.J.}\ }\textbf {\bibinfo {volume} {C58}},\ \bibinfo
  {pages} {639} (\bibinfo {year} {2008})},\ \Eprint
  {http://arxiv.org/abs/0803.0883} {arXiv:0803.0883 [hep-ph]} \BibitemShut
  {NoStop}%
%%CITATION = ARXIV:0803.0883;%%
\bibitem [{\citenamefont {Gieseke}\ \emph {et~al.}(2011)\citenamefont
  {Gieseke}, \citenamefont {Grellscheid}, \citenamefont {Hamilton},
  \citenamefont {Papaefstathiou}, \citenamefont {Platzer} \emph
  {et~al.}}]{Gieseke:2011na}%
  \BibitemOpen
  \bibfield  {author} {\bibinfo {author} {\bibfnamefont {S.}~\bibnamefont
  {Gieseke}}, \bibinfo {author} {\bibfnamefont {D.}~\bibnamefont
  {Grellscheid}}, \bibinfo {author} {\bibfnamefont {K.}~\bibnamefont
  {Hamilton}}, \bibinfo {author} {\bibfnamefont {A.}~\bibnamefont
  {Papaefstathiou}}, \bibinfo {author} {\bibfnamefont {S.}~\bibnamefont
  {Platzer}},  \emph {et~al.},\ }\href@noop {} {\  (\bibinfo {year} {2011})},\
  \Eprint {http://arxiv.org/abs/1102.1672} {arXiv:1102.1672 [hep-ph]}
  \BibitemShut {NoStop}%
%%CITATION = ARXIV:1102.1672;%%
\bibitem [{\citenamefont {Arnold}\ \emph {et~al.}(2012)\citenamefont {Arnold},
  \citenamefont {d'Errico}, \citenamefont {Gieseke}, \citenamefont
  {Grellscheid}, \citenamefont {Hamilton} \emph {et~al.}}]{Arnold:2012fq}%
  \BibitemOpen
  \bibfield  {author} {\bibinfo {author} {\bibfnamefont {K.}~\bibnamefont
  {Arnold}}, \bibinfo {author} {\bibfnamefont {L.}~\bibnamefont {d'Errico}},
  \bibinfo {author} {\bibfnamefont {S.}~\bibnamefont {Gieseke}}, \bibinfo
  {author} {\bibfnamefont {D.}~\bibnamefont {Grellscheid}}, \bibinfo {author}
  {\bibfnamefont {K.}~\bibnamefont {Hamilton}},  \emph {et~al.},\ }\href@noop
  {} {\  (\bibinfo {year} {2012})},\ \Eprint {http://arxiv.org/abs/1205.4902}
  {arXiv:1205.4902 [hep-ph]} \BibitemShut {NoStop}%
%%CITATION = ARXIV:1205.4902;%%
\bibitem [{\citenamefont {Bellm}\ \emph {et~al.}(2013)\citenamefont {Bellm},
  \citenamefont {Gieseke}, \citenamefont {Grellscheid}, \citenamefont
  {Papaefstathiou}, \citenamefont {Platzer} \emph {et~al.}}]{Bellm:2013lba}%
  \BibitemOpen
  \bibfield  {author} {\bibinfo {author} {\bibfnamefont {J.}~\bibnamefont
  {Bellm}}, \bibinfo {author} {\bibfnamefont {S.}~\bibnamefont {Gieseke}},
  \bibinfo {author} {\bibfnamefont {D.}~\bibnamefont {Grellscheid}}, \bibinfo
  {author} {\bibfnamefont {A.}~\bibnamefont {Papaefstathiou}}, \bibinfo
  {author} {\bibfnamefont {S.}~\bibnamefont {Platzer}},  \emph {et~al.},\
  }\href@noop {} {\  (\bibinfo {year} {2013})},\ \Eprint
  {http://arxiv.org/abs/1310.6877} {arXiv:1310.6877 [hep-ph]} \BibitemShut
  {NoStop}%
%%CITATION = ARXIV:1310.6877;%%
\bibitem [{\citenamefont {Cascioli}\ \emph {et~al.}(2012)\citenamefont
  {Cascioli}, \citenamefont {Maierhoefer},\ and\ \citenamefont
  {Pozzorini}}]{Cascioli:2011va}%
  \BibitemOpen
  \bibfield  {author} {\bibinfo {author} {\bibfnamefont {F.}~\bibnamefont
  {Cascioli}}, \bibinfo {author} {\bibfnamefont {P.}~\bibnamefont
  {Maierhoefer}}, \ and\ \bibinfo {author} {\bibfnamefont {S.}~\bibnamefont
  {Pozzorini}},\ }\href {\doibase 10.1103/PhysRevLett.108.111601} {\bibfield
  {journal} {\bibinfo  {journal} {Phys.Rev.Lett.}\ }\textbf {\bibinfo {volume}
  {108}},\ \bibinfo {pages} {111601} (\bibinfo {year} {2012})},\ \Eprint
  {http://arxiv.org/abs/1111.5206} {arXiv:1111.5206 [hep-ph]} \BibitemShut
  {NoStop}%
%%CITATION = ARXIV:1111.5206;%%
\bibitem [{\citenamefont {Frixione}\ \emph {et~al.}(2011)\citenamefont
  {Frixione}, \citenamefont {Stoeckli}, \citenamefont {Torrielli},\ and\
  \citenamefont {Webber}}]{Frixione:2010ra}%
  \BibitemOpen
  \bibfield  {author} {\bibinfo {author} {\bibfnamefont {S.}~\bibnamefont
  {Frixione}}, \bibinfo {author} {\bibfnamefont {F.}~\bibnamefont {Stoeckli}},
  \bibinfo {author} {\bibfnamefont {P.}~\bibnamefont {Torrielli}}, \ and\
  \bibinfo {author} {\bibfnamefont {B.~R.}\ \bibnamefont {Webber}},\ }\href
  {\doibase 10.1007/JHEP01(2011)053} {\bibfield  {journal} {\bibinfo  {journal}
  {JHEP}\ }\textbf {\bibinfo {volume} {1101}},\ \bibinfo {pages} {053}
  (\bibinfo {year} {2011})},\ \Eprint {http://arxiv.org/abs/1010.0568}
  {arXiv:1010.0568 [hep-ph]} \BibitemShut {NoStop}%
%%CITATION = ARXIV:1010.0568;%%
\bibitem [{\citenamefont {Frederix}\ \emph {et~al.}(2011)\citenamefont
  {Frederix}, \citenamefont {Frixione}, \citenamefont {Hirschi}, \citenamefont
  {Maltoni}, \citenamefont {Pittau} \emph {et~al.}}]{Frederix:2011zi}%
  \BibitemOpen
  \bibfield  {author} {\bibinfo {author} {\bibfnamefont {R.}~\bibnamefont
  {Frederix}}, \bibinfo {author} {\bibfnamefont {S.}~\bibnamefont {Frixione}},
  \bibinfo {author} {\bibfnamefont {V.}~\bibnamefont {Hirschi}}, \bibinfo
  {author} {\bibfnamefont {F.}~\bibnamefont {Maltoni}}, \bibinfo {author}
  {\bibfnamefont {R.}~\bibnamefont {Pittau}},  \emph {et~al.},\ }\href
  {\doibase 10.1016/j.physletb.2011.06.012} {\bibfield  {journal} {\bibinfo
  {journal} {Phys.Lett.}\ }\textbf {\bibinfo {volume} {B701}},\ \bibinfo
  {pages} {427} (\bibinfo {year} {2011})},\ \Eprint
  {http://arxiv.org/abs/1104.5613} {arXiv:1104.5613 [hep-ph]} \BibitemShut
  {NoStop}%
%%CITATION = ARXIV:1104.5613;%%
\bibitem [{\citenamefont {Alwall}\ \emph {et~al.}(2014)\citenamefont {Alwall},
  \citenamefont {Frederix}, \citenamefont {Frixione}, \citenamefont {Hirschi},
  \citenamefont {Maltoni} \emph {et~al.}}]{Alwall:2014hca}%
  \BibitemOpen
  \bibfield  {author} {\bibinfo {author} {\bibfnamefont {J.}~\bibnamefont
  {Alwall}}, \bibinfo {author} {\bibfnamefont {R.}~\bibnamefont {Frederix}},
  \bibinfo {author} {\bibfnamefont {S.}~\bibnamefont {Frixione}}, \bibinfo
  {author} {\bibfnamefont {V.}~\bibnamefont {Hirschi}}, \bibinfo {author}
  {\bibfnamefont {F.}~\bibnamefont {Maltoni}},  \emph {et~al.},\ }\href
  {\doibase 10.1007/JHEP07(2014)079} {\bibfield  {journal} {\bibinfo  {journal}
  {JHEP}\ }\textbf {\bibinfo {volume} {1407}},\ \bibinfo {pages} {079}
  (\bibinfo {year} {2014})},\ \Eprint {http://arxiv.org/abs/1405.0301}
  {arXiv:1405.0301 [hep-ph]} \BibitemShut {NoStop}%
%%CITATION = ARXIV:1405.0301;%%
\bibitem [{\citenamefont {Lai}\ \emph {et~al.}(2010)\citenamefont {Lai},
  \citenamefont {Guzzi}, \citenamefont {Huston}, \citenamefont {Li},
  \citenamefont {Nadolsky} \emph {et~al.}}]{Lai:2010vv}%
  \BibitemOpen
  \bibfield  {author} {\bibinfo {author} {\bibfnamefont {H.-L.}\ \bibnamefont
  {Lai}}, \bibinfo {author} {\bibfnamefont {M.}~\bibnamefont {Guzzi}}, \bibinfo
  {author} {\bibfnamefont {J.}~\bibnamefont {Huston}}, \bibinfo {author}
  {\bibfnamefont {Z.}~\bibnamefont {Li}}, \bibinfo {author} {\bibfnamefont
  {P.~M.}\ \bibnamefont {Nadolsky}},  \emph {et~al.},\ }\href {\doibase
  10.1103/PhysRevD.82.074024} {\bibfield  {journal} {\bibinfo  {journal}
  {Phys.Rev.}\ }\textbf {\bibinfo {volume} {D82}},\ \bibinfo {pages} {074024}
  (\bibinfo {year} {2010})},\ \Eprint {http://arxiv.org/abs/1007.2241}
  {arXiv:1007.2241 [hep-ph]} \BibitemShut {NoStop}%
%%CITATION = ARXIV:1007.2241;%%
\end{thebibliography}%
\bibliographystyle{apsrev4-1}

\end{document}